\documentstyle[12pt,epsfig,here]{article}
\def\lsim{\mathrel{\rlap{\lower3pt\hbox{\hskip0pt$\sim$}}
    \raise1pt\hbox{$<$}}}
\def\gsim{\mathrel{\rlap{\lower4pt\hbox{\hskip1pt$\sim$}}
    \raise1pt\hbox{$>$}}}
\begin{document}
\newcommand{\beq}{\begin{equation}}
\newcommand{\eeq}{\end{equation}}
\def\beqn{\begin{eqnarray}}
\def\eeqn{\end{eqnarray}}
\newcommand{\qt}{\tilde q}
\newcommand{\Tr}{{\rm Tr}\,}
\newcommand{\E}{{\cal E}}
\newcommand{\qtu}{\tilde q_{1}}
\newcommand{\qtd}{\tilde q_{2}}
\newcommand{\ntwo}{${\cal N}=2\;$}
\newcommand{\none}{${\cal N}=1\;$}
\newcommand{\vp}{\varphi}
\newcommand{\pt}{\partial}
\renewcommand{\theequation}{\thesection.\arabic{equation}}


\begin{titlepage}
\renewcommand{\thefootnote}{\fnsymbol{footnote}}

\begin{flushright}
FTPI-MINN-04/07, UMN-TH-2234/04\\
ITEP-TH-12/04
\end{flushright}


\vfil

\begin{center}
\baselineskip20pt
{ \Large \bf   Non-Abelian String Junctions
as Confined Monopoles}
\end{center}
\vfil
\begin{center}

\vspace{0.3cm}

{\large
{ \bf    M.~Shifman$^{a}$} and { \bf A.~Yung$^{a,b,c}$}}

\vspace{0.3cm}

$^a${\it  William I. Fine Theoretical Physics Institute,
University of Minnesota,
Minneapolis, MN 55455, USA}\\
$^b${\it Petersburg Nuclear Physics Institute, Gatchina, St. Petersburg
188300, Russia}\\
$^c${\it Institute of   Theoretical and Experimental Physics,
Moscow  117250, Russia}\\

\vfil

{\large\bf Abstract}
\end{center}
\vspace*{.25cm}
\vfil

Various dynamical regimes associated
with confined monopoles in the Higgs phase
of ${\cal N}=2$ two-flavor QCD are studied. The microscopic
model we deal with has the  SU(2)$\times$U(1) gauge group, with a
Fayet-Iliopoulos term of the U(1) factor,
and large and (nearly) degenerate mass terms of the
matter  hypermultiplets. We present a complete quasiclassical treatment
of the BPS sector of this model, including the full set of the
first-order equations, derivations of all relevant zero modes,
and derivation of an effective low-energy theory for
the corresponding collective coordinates. The macroscopic description is
provided by
a $CP^1$ model with or without twisted mass. The confined monopoles --
string junctions of the microscopic theory -- are mapped onto BPS kinks
of the $CP^1$ model. The string junction is
1/4 BPS. Masses and other characteristics of
the confined monopoles are matched with those of the
$CP^1$-model kinks. The matching demonstrates
the occurrence of an anomaly in the monopole central charge
in 4D Yang-Mills theory. We study what becomes of the confined
monopole in the {\em bona fide} non-Abelian limit of degenerate mass terms
where a global SU(2) symmetry is restored.
The solution of the macroscopic
model is known  e.g. from the mirror description of the $CP^1$ model.
The monopoles, aka $CP^1$-model kinks, are stabilized by nonperturbative
dynamics of  the  $CP^1$ model. We explain an earlier
rather puzzling observation
of a correspondence  between the BPS kink spectrum in the $CP^1$ model
and the Seiberg-Witten solution.

\end{titlepage}

\newpage

\tableofcontents

\newpage

\section{Introduction}
\label{introduction}
\renewcommand{\theequation}{\thesection.\arabic{equation}}
\setcounter{equation}{0}

This is the third work in  the project devoted to
investigation of the string/D-brane phenomena in supersymmetric gauge
theories \cite{Shifman:2002jm,SYsu3wall}. In the first work
\cite{Shifman:2002jm}
we studied Abelian strings ending on BPS
domain walls (D-branes) and localization of an (2+1)-dimensional
U(1) gauge field on the wall. The second work \cite{SYsu3wall}
was devoted to localization of non-Abelian fields on a stack of BPS
domain walls, and 1/4 BPS junctions of non-Abelian strings with the
walls. Here we extend the analysis and study 1/4 BPS non-Abelian string
junctions. This very interesting phenomenon has a clear-cut physical
picture behind
it --- it  describes monopoles in the confined phase. We are building
the present
analysis on our previous results, as well as on the results of
Refs.~\cite{Auzzi:2003fs,Dorey,Hanany:2003hp,Tong:2003pz} interspersed
in the fabric of the present work.

Results and techniques
of string/D-brane theory, being applied to non-Abelian field
theories (both, supersymmetric and non-supersymmetric),
lead to qualitative and quantitative predictions which became especially
numerous after the discovery \cite{AdSCFT} of the ADS/CFT correspondence.
If gauge theories at strong coupling are in a sense dual to
string/D-brane theory, they must support domain walls
(or D-branes) \cite{P},
and we know, they do \cite{DS,Witten:1997ep}.
In addition, string/D-brane theory teaches us that a
fundamental string that starts on a confined quark, can end
on the domain wall. In the dual description the confined
quark becomes a confined monopole.  This is our primary object of study
in the present paper. The question we ask is how the confined monopole
connects to the flux tube.

Our task is to study this phenomenon in a controllable manner.
To this end we will use a unique model
which emerged recently as {\em the one} providing
an ideal theoretical environment. The model will allows us  to
study, in a quasiclassical regime,
confined monopoles whose magnetic flux flows
through non-Abelian strings attached to them.
The predictive power of the model derives from the fact
that it has {\em exact}  ${\cal N}=2$ supersymmetry
(i.e. eight supercharges).
The non-Abelian strings are 1/2 BPS, while the confined monopoles
are 1/4 BPS saturated.

Let us outline some basic features
of this model, which we will refer to as ``microscopic."
We consider \ntwo QCD  \cite{SW1,SW2}
with the gauge group SU(2)$\times$U(1) with $N_f=2$ flavors of massive
fundamental matter hypermultiplets (quarks)
perturbed by the Fayet-Iliopoulos (FI) term \cite{FI} of the U(1)
factor.  In this theory we focus on a special so-called $r=2$ vacuum
\cite{APS,CKM,MY} in which two quark flavors develop
vacuum expectation values (VEV's).
This vacuum is at weak coupling  if  the quark mass terms $m_{1,2}$
are large enough. They may or may not be equal.
We consider both cases. If  $m_1=m_2$ the SU(2) gauge
group remains unbroken by VEV's of the adjoint fields. The quark
condensation does
break the gauge group SU(2)$\times$U(1) at a scale $\xi$
(FI parameter) but  leaves
a global diagonal SU(2)$_{C+F}$ subgroup of the gauge and flavor groups
unbroken. It was recently shown  \cite{Auzzi:2003fs} that in this case
the flux tubes
(strings) acquire additional orientational zero modes associated
with rotation of the color magnetic flux inside the SU(2)
group (similar results in three dimensions were
obtained in \cite{Hanany:2003hp}).  This makes them genuinely non-Abelian.
Moreover, it was found \cite{Auzzi:2003fs,Hanany:2003hp} that
the low-energy
dynamics of  the orientational zero modes of the non-Abelian strings
are described by an effective (1+1)-dimensional $CP^1$
model\,\footnote{$CP^{N-1}$  for the gauge group SU($N$)$\times$U(1).}
on the string
world sheet. The two-dimensional $CP^1$ model describing dynamics
of the collective coordinates will be referred to as ``macroscopic."

In fact, we can view our SU(2)$\times$U(1)
microscopic theory as a theory with the gauge group
SU(3) broken down to SU(2)$\times$U(1)  at a large scale, of the order
of $m_{1,2}$. This SU(3) theory has three types of monopoles associated
with three roots of SU(3) algebra. Two of them are confined by
``elementary" strings which we denote as (1,0) and (0,1)
\cite{MY} (here $(n,k)$ denotes the string with two winding
numbers $n$ and $k$
with respect to two U(1) subgroups of SU(3), see
Sect.~\ref{elementarystrings}  for more details).
These monopoles are very heavy, with masses of the order of
$m_{1,2}/g^2$,
and we do not touch  them in this paper. They were considered recently
in Ref.~\cite{ABEK}.  We will study only the monopoles which lie entirely
inside the SU(2) factor of the SU(3) ``proto"group. They are much
lighter than $m_{1,2}/g^2$.
Classically, on the Coulomb branch (i.e. when  the FI parameter $\xi$
vanishes),
their mass is proportional to $| \Delta m \,| /g^2$ where $\Delta m
=m_1-m_2$. In  the   limit $\Delta m\to 0$ they
become massless, formally, in the classical
approximation. Simultaneously their size become infinite \cite{We}.
The mass and size are stabilized by confinement effects
which are highly quantum. The confinement of monopoles occurs
on the Higgs branch, at $\xi\neq 0$. An interplay between
$  \Delta m$,  $\xi$ and the dynamical Yang-Mills scale $\Lambda$
leads\,\footnote{$\Lambda$ is also the dynamical scale of
the 2D $CP^1$ model.} to a spectrum of   dynamical scenarios,
all of which are interesting
and will be discussed in the present paper from a unified  point of view.

\begin{figure}[h]
\epsfxsize=12cm
\centerline{\epsfbox{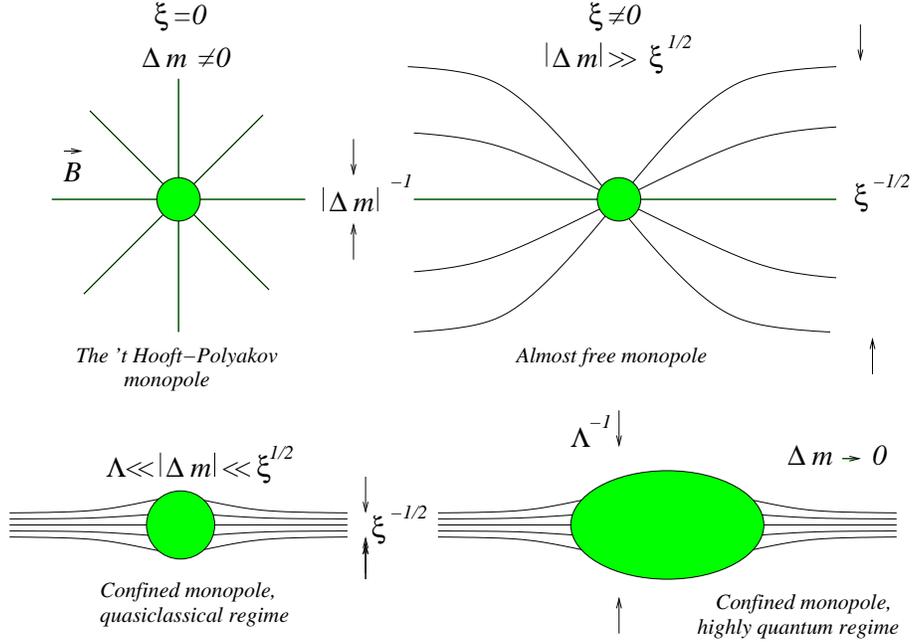}}
\caption{
Various regimes for the monopoles and
flux tubes. The dynamical scale parameters $\Lambda$
are the same in the microscopic and microscopic theories.
The effective world-sheet theory we derive -- the twisted-mass
$CP^1$ model -- applies
to the last two regimes: $\Lambda\ll|\Delta m |\ll \xi^{1/2}$ (the left
lower corner)
and $\Delta m \to 0$ (the right lower corner). The latter case
corresponds to the vanishing   twisted mass.}
\label{twoabcd}
\end{figure}

A qualitative evolution of the monopoles under consideration
as a function of the  relevant parameters is presented in
Fig.~\ref{twoabcd}.
We begin with the limit $\xi \to 0$ while $\Delta m $ is kept fixed.
Then the corresponding  microscopic theory supports
the conventional (unconfined)
't~Hooft-Polyakov monopoles \cite{thopo}
due to the spontaneous breaking of the {\em gauge} SU(2)
down to U(1),
\beq
\langle a^3\rangle  =-\frac{1}{\sqrt{2}}\Delta m \,.
\eeq
(the  upper left corner of Fig.~\ref{twoabcd}).
If we allow $\xi$ be non-vanishing but
\beq
|\Delta m | \gg\sqrt{\xi}
\eeq
then the effect which comes into play first is the above
spontaneous breaking of the   gauge SU(2).
Further gauge symmetry breaking, due to $\xi\neq 0$,
which leads to complete Higgsing of the model and the string
formation (confinement of monopoles) is much weaker.
Thus,  we deal here with  the formation
of ``almost"  't~Hooft-Polyakov monopoles, with a typical size
$\sim \left| \Delta m\right| ^{-1}\,.$ Only at much larger distances,
$\sim \xi ^{-1/2}$, the charge condensation enters the game,
and forces the magnetic flux, rather than spreading evenly a l\'a
Coulomb, to form flux tubes (the  upper right corner of
Fig.~\ref{twoabcd}).  There will be two such flux tubes, with the distinct
orientation of the color-magnetic flux. The monopoles, albeit confined,
are weakly confined.

Now, if we further reduce $\left| \Delta m\right| $,
\beq
\Lambda \ll \left| \Delta m\right|   \ll \sqrt{\xi}\, ,
\label{ququr}
\eeq
the size of the monopole ($\sim \left| \Delta m\right|^{-1} $) becomes
larger than the transverse size of the attached strings.
The monopole gets squeezed  in earnest by
the strings --- it becomes  a {\em bona fide} confined
monopole (the  lower left corner of  Fig.~\ref{twoabcd}).
A macroscopic description of such monopoles is provided
by  the twisted-mass $CP^1$ model, on which we will dwell below.
The value of the twisted mass $\mu = \Delta m$.
The confined monopole is nothing but
the twisted-mass sigma-model kink
which has a typical size $\sim \left| \mu\right|^{-1} $.

As we further diminish $\left| \Delta m\right|$
approaching $\Lambda$ and then getting  below $\Lambda$,
the size of the monopole grows, and, classically, it would explode.
This is where quantum effects in the world-sheet theory take over.
It is natural to refer to this domain of parameters as the ``regime of
highly quantum dynamics."
While the thickness of the string (in the transverse direction) is
$\sim \xi ^{-1/2}$, the
$z$-direction size of the kink  representing the confined
monopole in the highly quantum regime is much larger, $\sim \Lambda^{-1}$,
see the  lower right corner of  Fig.~\ref{twoabcd}.

\begin{figure}[h]
\epsfxsize=7cm
\centerline{\epsfbox{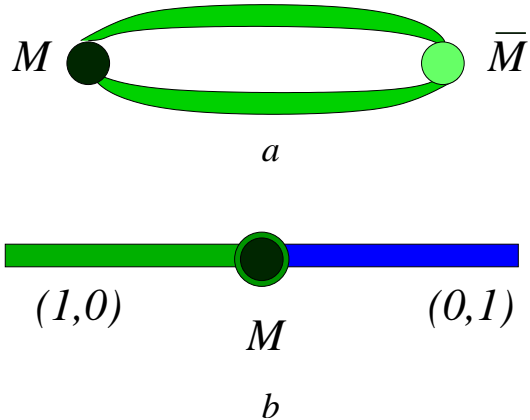}}
\caption{
(a) The monopole-antimonopole pair; (b) the  monopole with two
(infinitely long) elementary flux tubes attached to it.}
\label{oneab}
\end{figure}

While the monopoles on the Coulomb branch
(i.e. those of the 't Hooft-Polyakov type) are thoroughly discussed
in  the literature, the Higgs-branch monopoles -- the confined monopoles --
received much less attention. We intend to close this gap.
We will study what becomes of the
non-Abelian SU(2) monopole in the confinement
phase which is set by the quark condensation and formation of
the flux tubes (at non-vanishing FI parameter $\xi\neq 0$).
The monopole-antimonopole pair will be   confined in the meson-like
state by a composite string,  a bound state of two
``elementary" strings \cite{MY}, see Fig. ~\ref{oneab}a. Unfortunately, such
mesons are  unstable and cannot be studied in the static limit.
Instead, we will focus on another (static) field configuration
typical of the confinement phase:
an SU(2) monopole with two semi-infinite elementary strings attached to it,
see Fig. ~\ref{oneab}b. This configuration, a junction of two  elementary
strings,
 is stable and, moreover, 1/4-BPS saturated.

We perform a complete quasiclassical analysis of our microscopic model.
A full set of the first-order master equations describing all 1/2 and 1/4
BPS topological
defects -- domain walls, strings, and all possible junctions -- is derived.
While the domain walls, strings and the wall-string
junctions were studied previously,
we expand the analysis to cover the case of the string junctions.
We find appropriate solutions of the master equations.
We derive zero modes, both bosonic and fermionic,
specific for non-Abelian strings and their junctions. The corresponding
collective coordinates (moduli) are introduced, which allows us to obtain
the macroscopic description of the topological
defects in question. We present solid quantitative evidence
that at  $\sqrt\xi\gg |\Delta m |$ the effective (1+1)-dimensional theory
on the string world sheet is the $CP^1$ model, and  that
the confined monopoles of the microscopic theory must be
identified with the  kinks of the macroscopic  $CP^1$ model.
In particular, we explicitly demonstrate
that the  first-order BPS equations of the 4D Yang-Mills theory
reduce to  first-order BPS equations for the $CP^1$-model kink.
We match the mass  of the confined 4D Yang-Mills monopole
with that of the $CP^1$-model kink.
This evidence completes the proof of the statements
that were made in the literature previously \cite{Tong:2003pz,SYsu3wall}.

The mass match mentioned above requires
the presence of an anomaly in the monopole central charge
of  4D Yang-Mills theory which must match its counterpart in the
$CP^1$ model. While the anomaly in the  $CP^1$ model
was known previously \cite{Losev}, that in the Yang-Mills theory
was not known. A fermion operator, an anomaly in the anticommutator
$\{Q\,, \, Q\}$, crucial for the confined monopoles,
was identified in our recent publication \cite{SYsu3wall}.
Here we further elaborate on this issue. Meanwhile,
another anomalous contribution, crucial for the monopoles in the
Coulomb regime, was identified and 
analyzed in Ref.~\cite{Rebhan:2004vn}.

The identification of the  confined monopoles of 4D Yang-Mills theory
with the  kinks of the 2D  $CP^1$ model gives us two advantages.
First, we can and do explore the highly quantum regime of
$\Delta m ,\,\, \mu \to 0$. In  this limit no quasiclassical treatment is
available. The global SU(2) gets restored. Confined monopoles do not disappear, they survive. Appropriate exploration tools
are available in the framework of the $CP^1$ model.

Second, we can and do explain a long-standing puzzling observation made
in Ref.~\cite{Dorey}.  A  comparison of  the corresponding central charges
revealed \cite{Dorey} a close parallel between
four-dimensional Yang-Mills theory with $N_f=2$ and the two-dimensional
$CP^1$ model. The observation referred to the Coulomb branch
of the Seiberg-Witten theory, with unconfined 't Hooft-Polyakov-like
monopoles/dyons. We clarify physics responsible for this correspondence.
In fact, the twisted-mass $CP^1$ model is equivalent
to the {\em Higgs phase} of four-dimensional Yang-Mills theory with
$N_f=2$.  These two theories are ``microscopic-macroscopic" partners.
We show, however, that the BPS data are independent
of the value of the FI parameter $\xi$, because this parameter has
the $R$ parity that does not match that of the central charges.
Therefore, in the BPS sector one can vary  the FI parameter at will;
in particular, pass to the limit $\xi\to 0$, where one finds oneself on
the Coulomb branch. Thus, the {\em bona fide}
correspondence refers to the Higgs phase where it is rather obvious,
but  holomorphy allows one to extend it to the Coulomb branch too.

The paper is organized as follows. In Sect.~\ref{abriefsummary}  we outline
our microscopic 4D Yang-Mills theory: SU(2)$\times$U(1) two-flavor QCD with the Fayet-Iliopoulos term and extended ${\cal N}=2$
supersymmetry. Section \ref{nonabelianstrings} is devoted to
description of the non-Abelian strings
and introduction of orientational moduli.
In Sect.~\ref{macroscopictheory} we derive and study the effective 2D
theory on the string world sheet. In particular, we show,
by virtue of an explicit calculation, that a
non-vanishing $\Delta m$ induces the twisted mass term $\mu$ in the
world-sheet
$CP^1$ model, and $\Delta m =\mu$.
Next we calculate superorientational (fermion) zero modes
of the non-Abelian string. In Sect.~\ref{sigmamodelkinks}
we establish and exploit the correspondence between the
monopole and the
junction of two  elementary  strings. We solve the first-order master
equations for the junction, match the monopole and kink masses,
discuss the multiplicity matching and other
implications for the confined monopoles.
In Sect.~\ref{quantumlimit} we consider monopoles/kinks
in the quantum limit $\Delta m,\,\,\mu\to 0$ and match  anomalies in
the central charges of the corresponding superalgebras.
In Sect.~\ref{dopo6}  we explain why the BPS sector of two-dimensional
$CP^1$ model is related to the Seiberg-Witten solution of 4D
super-Yang-Mills
theory: a direct correspondence is between $CP^1$ and the Higgs phase
of Yang-Mills; holomorphy of the central charges makes possible
a subsequent
transition to the Coulomb branch (the Seiberg-Witten solution).
Finally, Section \ref{conclusions} briefly summarizes our conclusions.

\section{A brief summary of the theoretical set-up}
\label{abriefsummary}
\renewcommand{\theequation}{\thesection.\arabic{equation}}
\setcounter{equation}{0}

Our task is to study the string/D-brane phenomena in supersymmetric
gauge theories in a fully controllable mode. An appropriate theoretical
set-up gradually emerged in the last three years
\cite{MY,Shifman:2002jm,Auzzi:2003fs,SYsu3wall} --- a particular
${\cal N}=2$ gauge model with judiciously chosen matter hypermultiplets
and a special adjustment of the matter mass terms. The model evolved
in the direction of simplification; currently it presents a
theoretical scene fully fit for
studies of the phenomena we are interested in.

The gauge symmetry of the model we will deal with is SU(2)$\times$U(1).
Besides the gauge bosons, gauginos and their ${\cal N}=2$ superpartners,
it has  a matter sector consisting of two ``quark" hypermultiplets,
with large (and {\em degenerate}  or {\em nearly degenerate}) mass terms.
One also  introduces a Fayet-Iliopoulos term, so that the overall
superpotential
takes the form
\begin{eqnarray}
{\cal W}& =&\frac{1}{\sqrt 2} \sum_{A=1}^2
\left( \tilde q_A {\cal A}  q^A +  \tilde q_A {\cal A}^a\,\tau^a
q^A\right)
\nonumber\\[3mm]
&+&  \sum_{A=1,2}m_A\, q^A   \tilde q_A -
\frac1{\sqrt{2}}\, \xi\, {\cal A}\,,
\label{sprptntl}
\end{eqnarray}
where ${\cal A}^a$ and ${\cal A}$ are  chiral superfields, the ${\cal N}=2$
 superpartners of the gauge bosons of SU(2)  and  U(1), respectively.
Furthermore,
 $q_A$ and $\tilde q_A$ ($A=1,2$) represent two matter hypermultiplets,
while $\xi$, $m_1$ and $m_2$ are constants, assumed to be much larger
than the dynamical scale parameter of the SU(2) gauge theory,
see Sect.~\ref{nonabelianstrings}.
The mass terms $m_{1,2}$ can always be made real and positive
by virtue of an appropriate field definition. We will assume this to be
the case. Moreover, for simplicity we will assume that
the Fayet-Iliopoulos parameter $\xi$
is real and positive too.\footnote{In fact, $\xi$ is
the first component of an SU(2)$_R$ vector $\vec\xi$ of the
generalized Fayet-Iliopoulos parameters introduced in
\cite{HSZ,VY}, see Eq.~(\ref{vyvector}) below.}
For further details of our theoretical set-up  the reader is
referred to Refs.~\cite{MY,Auzzi:2003fs,SYsu3wall}.
Besides local symmetries, (and besides a global SU(2)$_R$
inherent to ${\cal N}=2$), at $m_1=m_2$  the model has a global
SU(2) flavor symmetry associated  with rotations of the first
and second hypermultiplets. This ``symmetric" point is the focus
of the present work.
However, to make contact with  quasiclassical results, at times
we will leave the symmetric point  $m_1=m_2$  and consider a deformed
case $m_1\neq m_2$ assuming, however, the deformation to be small,
$|m_1-m_2|\ll m_{1,2}$. The reader is advised to exercise caution not
to confuse these two regimes.

The Fayet-Iliopoulos term triggers the spontaneous breaking
of the gauge symmetry. The vacuum expectation values (VEV's)
of the squark fields can be chosen as
\beqn
\langle q^{kA}\rangle &=&\langle \bar{\tilde{q}}^{kA}\rangle =\sqrt{
\frac{\xi}{2}}\, \left(
\begin{array}{cc}
1 & 0 \\
0 & 1\\
\end{array}
\right),
\nonumber\\[3mm]
k&=&1,2,\qquad A=1,2\,,
\label{qvev}
\eeqn
{\em up to gauge rotations}. The color-flavor locked form of VEV's
in Eq.~(\ref{qvev})
results in the fact that, while the theory is fully Higgsed, a diagonal
SU(2)$_{C+F}$ survives as a global symmetry (in the limit $m_1=m_2$).
This is a particular case  of the Bardakci-Halpern mechanism \cite{BarH}.
The most economic way to see the occurrence of the above  global
symmetry is through the matrix notation
 \beq
Q =\left(
\begin{array}{cc}
q^{11} & q^{12}\\[2mm]
q^{21} & q^{22}
\end{array}
\right)
\label{matrixphi}
\eeq
(and the same for $\tilde q$). Here $Q$ is a $2\times 2$ matrix, the first
superscript refers to SU(2) ``color" (we will also use the notation
$\varphi^r$ and  $\varphi^b$ meaning red and blue), while the second
($A=1$ or 2) to ``flavor." The covariant derivatives are defined in
such a way
that they act from the {\em left},
\beq
\nabla_\mu \, Q \equiv  \left( \partial_\mu -\frac{i}{2}\; A_{\mu}
-i A^{a}_{\mu}\, \frac{\tau^a}{2}\right)Q\, ,
\label{dcde}
\eeq
while the global flavor SU(2) transformations then act on $Q$ from the right.
Equation (\ref{dcde}) also shows our U(1) charge convention.
Needless to say,  ${\cal N}=2$ supersymmetry of the model is unaffected
by the gauge symmetry breaking (\ref{qvev}).  The local and global
symmetries
of the model and of the vacuum state are summarized in Table~\ref{table2}.
In the vacuum, there are no massless modes, all excitations
are massive.
\begin{table}
\begin{center}
\begin{tabular}{|c|c |}
\hline
${\cal N}=2$ SUSY  &  unbroken
  \\[3mm]
\hline
SU(2)$_R$  & unbroken
\\[2mm]
\hline
$\Delta m = 0:\,\, $\{U(1)$\times$ SU(2)\}$_G\times$ SU(2)
$\times$ U(1)
&${\rm U}(1)_{\rm diag}\times{\rm SU}(2)_{\rm diag}$
\\[2mm]
\hline
$\Delta m \neq 0:\,\, $\{U(1)$\times$ SU(2)\}$_G\times$  U(1)
$\times$ U(1)
&${\rm U}(1)_{\rm diag}$
\\
\hline
\end{tabular}
\end{center}
\caption{Symmetries of the microscopic theory and the pattern of the symmetry breaking
in the vacuum.}
\label{table2}
\end{table}

Concluding this section it would be in order to present
a broader perspective on our theoretical set-up. This will hopefully
provide additional conceptual insights  albeit technically this aspect
will not be pursued. We can view our model as a descendant of an
${\cal N}=2$  theory with the  SU(3) gauge group broken
down to SU(2)$\times$U(1) at a high scale,
of the order of the mass parameters $m_{1,2}$. This SU(3)
``proto"theory has three types of monopoles associated
with three roots of the SU(3) algebra. Two of them are confined by
``elementary" strings which we denote as (1,0) and (0,1) \cite{MY}.
(In general,  $(n,k)$ denotes the string with two winding numbers
$n$ and $k$ with respect to two U(1) Cartan subgroups of SU(3),
see  Sect.~\ref{nonabelianstrings} for  details.)
These monopoles are very heavy --- their masses are of the order of
$(m_{1,2})/g^2$ --- and we do not discuss them in this paper.  Note, however, that
they were considered recently in Ref.~\cite{ABEK}.
Our primary interest   is the ``third" monopole which lies entirely
inside the SU(2) factor of the SU(3) gauge group. It is much lighter
than the two mentioned above. This is the reason why in our studies
we settle for the SU(2)$\times$U(1) model while
the full SU(3) model is behind the scene.

\section{Non-Abelian strings}
\label{nonabelianstrings}
\setcounter{equation}{0}

In this section we review the formalism of Ref.~\cite{Auzzi:2003fs}
where non-Abelian strings were first introduced,
and make adjustments necessary for the present work.
We start from  U(1) $\times$U(1) moduli-free string solutions
found  in Ref.~ \cite{MY} in the case $m_1\neq m_2$.   Then we show
how additional orientational  zero modes arise in the limit
$m_1\to m_2$  making the strings at hand non-Abelian \cite{Auzzi:2003fs}.

Let us start from the case $m_1\neq m_2$.
One can readily convince oneself that as far as the  flux-tube
solutions are concerned  it is sufficient to limit oneself
to the following {\em ansatz} for  the matter fields:
\beq
q^{kA}=\bar{\qt}^{kA}\equiv
\frac{1}{\sqrt{2}}\,\varphi^{kA}\, .
\label{qqt}
\eeq
Correspondingly, we introduce the matrix
\beq
\Phi =\left(
\begin{array}{cc}
\varphi^{11} & \varphi^{12}\\[2mm]
\varphi^{21} & \varphi^{22}
\end{array}
\right)\,,\qquad \Phi =\sqrt 2\, Q =\sqrt 2\, \bar{\tilde Q}\,,
\label{phima}
\eeq
where the first
superscript refers to SU(2)  color , while the second  to flavor.
Note that the field identification (\ref{qqt}) is inappropriate
in dealing with quantum corrections, and, in particular, in the
zero-mode analysis.

Then the  bosonic part of the effective action of the model at hand
takes the
form\,\footnote{Here and below we use a formally  Euclidean notation, e.g.
$F_{\mu\nu}^2 = 2F_{0i}^2 + F_{ij}^2$,
$\, (\partial_\mu a)^2 = (\partial_0 a)^2 +(\partial_i a)^2$, etc.
This is appropriate since we are  going to study static (time-independent)
field configurations, and $A_0 =0$. Then the Euclidean action is
nothing but the energy functional. Furthermore, we
define $\sigma^{\alpha\dot{\alpha}}=(1,-i\vec{\tau})$,
 $\bar{\sigma}_{\dot{\alpha}\alpha}=(1,i\vec{\tau})$. Lowing and raising
of spinor indices is performed by  virtue of the antisymmetric tensor
defined as $\varepsilon_{12}=\varepsilon_{\dot{1}\dot{2}}=1$,
 $\varepsilon^{12}=\varepsilon^{\dot{1}\dot{2}}=-1$.
The same raising and lowering convention applies to the flavor SU(2)
indices $f$, $g$, etc. }
\beqn
S &=& \int {\rm d}^4x\left\{\frac1{4g_2^2}
\left(F^{a}_{\mu\nu}\right)^{2}
+ \frac1{4g_1^2}\left(F_{\mu\nu}\right)^{2}
+\frac1{g_1^2} |\pt_{\mu} a|^2 + \frac1{g_2^2} |D_{\mu} a^a|^2
 \right.
 \nonumber\\[3mm]
&+&
 {\rm Tr}\, (\nabla_\mu \Phi)^\dagger \,(\nabla_\mu \Phi )
+\frac{g^2_2}{8}\left[{\rm Tr}\,
\left(\Phi^\dagger \tau^a \Phi\right)\right]^2
 +
 \frac{g^2_1}{8}\left[ {\rm Tr}\,
\left( \Phi^\dagger \Phi \right)-2\xi \right]^2
 \nonumber\\[4mm]
&+&
\left.
\frac12\,{\rm Tr}\, \left\{ \left(\Phi^\dagger  [a_a\tau^a+a]^\dagger
+\sqrt{2}M^\dagger \Phi^\dagger \right) \left([a_a\tau^a+a]\Phi
+\Phi\sqrt{2}M \right)  \right\}
 \right\} \,,
\label{redqed}
\eeqn
where $a$ is the lowest component of the chiral superfield ${\cal A}$.
Here we introduced a $2\times 2$ mass matrix  $M$
acting on the flavor indices of $\Phi$. With our convention $M$ is diagonal,
\beq
M =\left(
\begin{array}{cc}
m_1 & 0\\[2mm]
0 & m_2
\end{array}
\right)\,.
\label{mama}
\eeq
Equation
(\ref{redqed}) implies, that besides the squark VEV's (\ref{qvev}),
the field $a$  and $a^a$ develop  vacuum expectations value too,
\beq
a=-\sqrt 2 m\,,\qquad a^3 =-\frac1{\sqrt{2}}\Delta m\,,
\label{avev}
\eeq
where
\beq
m=\frac12\left( m_1+m_2\right)\,, \qquad \Delta m=m_1-m_2\,.
\label{mdeltam}
\eeq
We see that
at $m_1\ne m_2$ the SU(2)$\times$U(1) gauge symmetry is broken down
to U(1)$\times $U(1) by the VEV of the adjoint scalar field $a^3$.
For definiteness, we will assume that $m_1 > m_2$. Then $\Delta m$ is
positive.\footnote{To keep our notation concise  we
will also use the parameter
$\mu \equiv m_1-m_2\equiv \Delta m$. This parameter is not
to be confused with the ${\cal N}=2$ breaking perturbation,
e.g.  Refs.~\cite{Shifman:2002jm,SYsu3wall}, which is routinely denoted by
$\mu$ following Seiberg and Witten \cite{SW1,SW2}.
In most cases $\mu$ will be reserved for the twisted mass in the macroscopic $CP^1$ model, while the mass splitting
$\Delta m$ will be used in the microscopic 4D Yang-Mills theory. We will show that $\mu =\Delta m$.  In some sections (e.g.
Sects.~\ref{unequalquark} and \ref{quantumlimit}), for brevity,  $\mu$
will replace $\Delta m$ in the microscopic 4D Yang-Mills theory.}
If $| \Delta m| \gg \xi^{1/2}$ then the gauge symmetry
breaking by $\langle  a^3\rangle $ has a larger scale than that by
the squark fields. This is a more important effect leading to monopoles with
masses $\sim | \Delta m|/g_2^2$. Formation of strings is governed by
$\xi$ and can be viewed as a ``secondary" effect.

On the other hand, if $\Delta m = 0$, the gauge SU(2) group
is unbroken by the adjoint scalar VEV's, since they reduce to
\beq
a=-\sqrt 2 m\,,\qquad a^3 =0\, .
\label{avevsu2}
\eeq
With two matter hypermultiplets, the  SU(2) part of the gauge group
is asymptotically free,  implying generation of a dynamical scale $\Lambda$.
If descent to  $\Lambda$ were uninterrupted, the gauge coupling
$g_2^2$ would explode at this scale.
Moreover,  strong coupling effects in the SU(2) subsector at the
scale $\Lambda$ would break the  SU(2) subgroup through the
Seiberg-Witten mechanism \cite{SW1}.  Since we want to stay
at weak coupling   we assume
that $\sqrt{\xi}\gg \Lambda$,
so that the SU(2) coupling running is frozen by the squark condensation
at a small value
\beq
\frac{8\pi^2}{g_2^2}=2\ln{\frac{\sqrt{\xi}}{\Lambda}} +\cdots \gg 1\,.
\label{g2}
\eeq
 Alternatively one can say that
\beq
\Lambda^2 = \xi \exp\left(-\frac{8\pi^2}{g_2^2 (\xi)}
\right)
\label{alter}
\eeq
with $g_2^2 (\xi)\ll 1$.

\subsection{Abrikosov-Nielsen-Olesen  vs. elementary strings}
\label{elementarystrings}

\begin{figure}[h]
\epsfxsize=4cm
\centerline{\epsfbox{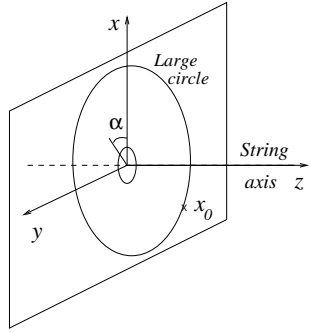}}
\caption{
Geometry of the string.}
\label{mmmon}
\end{figure}

To warm up, let us discuss the conventional Abrikosov-Nielsen-Olesen
(ANO) string \cite{ANO} in our model.\footnote{This subsection
is insensitive with regards to the choice of $\Delta m$ which may or
may not vanish.}
The existence of the ANO string is due to the fact that
$\pi_1({\rm U(1)}) = Z$, ensuring its topological stability.
For this solution one can discard the SU(2)$_c$ part of the action
($c$ stands for ``color"), putting $A_\mu^a=a^a\equiv 0$.
Correspondingly, there is no SU(2) winding of $\Phi$. Non-trivial topology
is realized through the U(1) winding of $\Phi$,
\beq
\Phi (x) =  \sqrt\xi \, e^{i\alpha (x)} \,,\qquad |x|\to \infty\,,
\label{frione}
\eeq
and
\beq
A_\ell = -2\, \varepsilon_{\ell k} \, \frac{x_k}{r}\,,\qquad \ell , k =1,2\,,
\label{fritwo}
\eeq
where $\alpha$ is the angle in the perpendicular
plane (Fig.~\ref{mmmon}), and $r$ is the distance from the string axis in
the perpendicular
plane. Equations (\ref{frione}) and  (\ref{fritwo})
refer to the minimal ANO string, with the minimal winding.
Needless to say, the tension of the ANO string is given by the standard
formula
\beq
T_{\rm ANO} =4\pi\xi\,.
\label{anof}
\eeq
This is not the string we are interested in here, however ---
in fact, in the problem at hand there are ``more elementary" strings
with 1/2
of the above tension, so that the ANO string can be viewed as
 a bound state of two elementary strings.

Where do they come from?
Since $\pi_1({\rm SU(2)})$ is trivial, at first sight it might seem
that there are no new options. This conclusion is wrong --- one can
combine the
$Z_2$ center of SU(2) with the element $-1\in$U(1)
to get a topologically stable string-like solution
possessing both windings, in SU(2) and U(1), of the following type
\begin{eqnarray}
\Phi (x) &=& \sqrt\xi \exp\left[ i \alpha (x) \frac{1\pm\tau^3}{2}\right]\,,
\qquad |x|\to\infty\,,\nonumber\\[3mm]
A_\ell &=& -\, \varepsilon_{\ell k} \, \frac{x_k}{r}\,,\qquad A_\ell^3 =
\mp  \varepsilon_{\ell k} \, \frac{x_k}{r}\,, \qquad \ell , k =1,2\,,
\label{newwi}
\end{eqnarray}
Correspondingly, the U(1) magnetic flux is twice smaller
than in the ANO case.
Since it is only the U(1) magnetic flux that enters the expression
for the appropriate central charge (see below), the tension
of the flux tube generated by the winding
(\ref{newwi}) is
\beq
T_\pm = 2\pi\,\xi\,.
\label{pmst}
\eeq
The $\pm$ subscript corresponds to two types of elementary strings
in which either only $\varphi^r$ or only $\varphi^b$
are topologically non-trivial.

We will refer to the  strings corresponding to the boundary
conditions (\ref{newwi}) as (1,0) and  (0,1) for the following reasons.
For the case of  non-equal quark masses the SU(2)$\times$U(1) group
is broken by the adjoint scalar VEV
to U(1)$\times$U(1). We have a lattice of strings
labeled by two integer numbers $(n,k)$ associated with the windings
with respect to two gauge  U(1)  groups which are linear combinations
of the two U(1)'s above which are natural in the SU(3) ``proto"theory. In
this terminology the ANO string is the sum (0,1)+(1,0)=(1,1), see  \cite{MY}
for further details.

\subsection{Embedding; first-order equations for the elementary \\
strings}
\label{embedding}

The charges of the $(n,k)$-strings can be plotted in the Cartan plane of
the  SU(3)  algebra of the ``proto"theory. We will use the convention of
labeling the flux of a given  string by the magnetic charge of the
monopole which
produces this flux and can be attached to its end. This is possible since
both
the string fluxes and the monopole charges are
elements of the group $\pi_1(U(1)^2) = {Z}^{2}$. This convention is
convenient because specifying the flux of a given string automatically fixes
the charge of the monopole that it confines.

\begin{figure}[h]
\epsfxsize=7cm
\centerline{\epsfbox{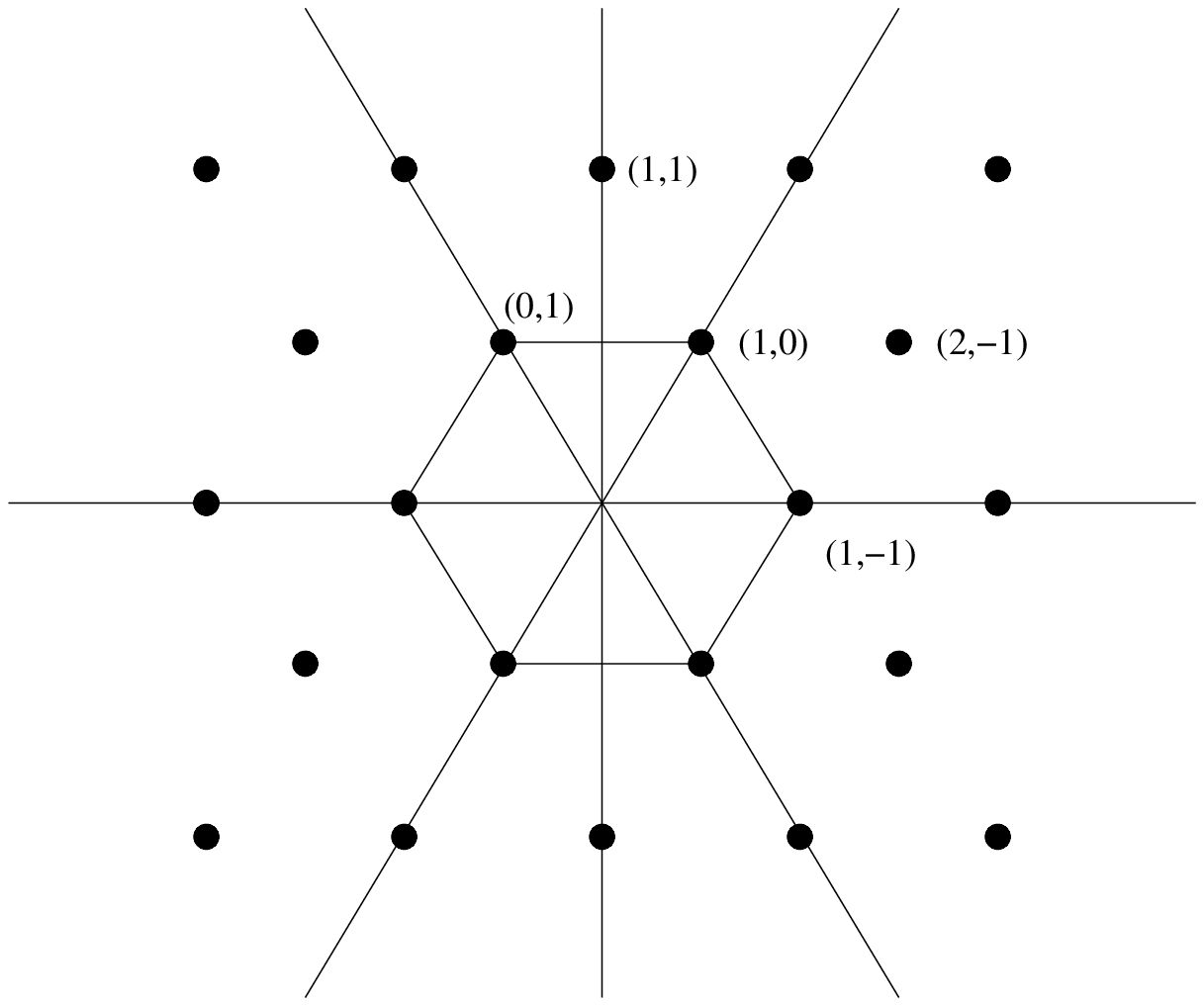}}
\caption{ Lattice of $(n,k)$   vortices.    }
\label{fi:lattice}
\end{figure}

Our strings are formed by the condensation of squarks
which have electric charges equal to the  weights of the SU(3)
algebra. The Dirac quantization condition tells us \cite{MY} that
the lattice of  the $(n,k)$-strings is  formed  by the roots of the  SU(3)
algebra.
This lattice of  the $(n,k)$-strings is shown in
Fig.~\ref{fi:lattice}. Two strings  $(1,0)$ and $(0,1)$ are the
``elementary'' or ``minimal"
BPS strings. All other strings can be considered as \ bound states of these
elementary  strings. If we plot two lines along the charges of these
 elementary  strings (see Fig.~\ref{fi:lattice}) they divide the
lattice into four sectors. It turns out \cite{MY} that the strings
in the upper and lower sectors are BPS but they are marginally
unstable. On the contrary, the strings lying in the right and left sectors
are (meta)stable bound states of the  elementary ones;  they are {\em not}
BPS-saturated.

Our study objective is the elementary string.
Since the both strings, (0,1) and (1,0), are 1/2 BPS-saturated,
Eq.~(\ref{pmst}) is
exact. These strings are {\em absolutely degenerate}. The degeneracy holds
beyond the classical level, with inclusion  of  quantum corrections,
perturbative  and nonperturbative. This is important for what follows.
The existence of two distinct strings with exactly degenerate tensions
is a special feature of supersymmetry implying the existence of a 2-string
junction.
A similar phenomenon for domain walls --- 2-wall junctions --- was
studied recently
\cite{asrptobe}.

The first-order equations for the BPS strings following
from the action (\ref{redqed}) are
\begin{eqnarray}
& &  F^{*a}_{3}+
     \frac{g^2_2}{2}\, \sigma\,
\left(\bar{\vp}_A\tau^a \vp^A\right)=0   , \qquad  a=1,2,3;
\nonumber \\[3mm]
& &   F^{*}_{3}+
     \frac{g^2_1}{2}\, \sigma\,
\left(|\vp^A|^2-2\xi \right)=0;
\nonumber \\[4mm]
 & &   ( \nabla_1  +i \, \sigma\,
\nabla_2)\, \vp^A=0, \qquad
\label{F38}
\end{eqnarray}
where $\sigma\, = \pm 1$ is the sign of the total flux and
\beq
F^*_m =\frac{1}{2}\, \varepsilon_{mnk}\, F_{nk}\,,\qquad m,n,k = 1,2,3\,.
\label{fstar}
\eeq
(see Ref.~\cite{Auzzi:2003fs} and Sect.~\ref{firstorder}).
To construct  the (0,1) and (1,0) strings
we further restrict  the gauge field $A_{\mu}^a$ to the single
color component $A_{\mu}^3$,
setting $A_{\mu}^1 = A_{\mu}^2=0 $, and consider only the squark
fields  of the   $2 \times 2$  color-flavor diagonal form,
\beq
\label{ansatz}
\vp^{kA}(x)   \ne 0,  \qquad  {\rm for}
\quad   k=A   =1,2.
\eeq
The off-diagonal  components of the matrix $\Phi$ are set to zero.

The (1,0) string arises when  the first flavor has the unit winding number
 while
the second flavor  does not wind at all. And {\em vice versa},  the
(0,1) string arises when  the second flavor has the unit winding number
while
the first flavor  does not wind. Consider for definiteness the (1,0)
string.  The solutions of the
first-order equations (\ref{F38}) will be sought for using the following
{\em ansatz} \cite{MY}:
\beqn
\Phi(x)
&=&
\left(
\begin{array}{cc}
  e^{ i \, \alpha  }\phi_1(r) & 0  \\
  0 &  \phi_2(r) \\
  \end{array}\right),
\nonumber\\[4mm]
A^3_{i}(x)
&=&
 -\varepsilon_{ij}\,\frac{x_j}{r^2}\
\left(1-f_3(r)\right),\;
\nonumber\\[4mm]
A_{i}(x)
&=&
 - \varepsilon_{ij}\,\frac{x_j}{r^2}\
\left(1-f(r)\right)\,
\label{sol}
\eeqn
where
the profile functions $\phi_1$, $\phi_2$ for the scalar fields and
$f_3$, $f$ for the gauge fields depend only on $r$ ($i,j=1,2$).
Applying this {\em ansatz} one can rearrange \cite{MY} the first-order
equations (\ref{F38}) in  the form
\beqn
&&
r\frac{d}{{d}r}\,\phi_1 (r)- \frac12\left( f(r)
+  f_3(r) \right)\phi_1 (r) = 0\, ,
\nonumber\\[4mm]
&&
r\frac{d}{{ d}r}\,\phi_2 (r)- \frac12\left(f(r)
-  f_3(r)\right)\phi_2 (r) = 0\, ,
\nonumber\\[4mm]
&&
-\frac1r\,\frac{ d}{{ d}r} f(r)+\frac{g^2_1}{2}\,
\left[\left(\phi_1(r)\right)^2 +\left(\phi_2(r)\right)^2-2\xi\right] =
0\, ,
\nonumber\\[4mm]
&&
-\frac1r\,\frac{d}{{ d}r} f_3(r)+\frac{g^2_2}{2}\,
\left[\left(\phi_1(r)\right)^2 -\left(\phi_2(r)\right)^2\right]  = 0
\, .
\label{foest}
\eeqn
Furthermore, one needs to specify the boundary conditions
which would determine the profile functions in these equations. Namely,
\beqn
&&
f_3(0) = 1\, ,\qquad f(0)=1\, ;
\nonumber\\[4mm]
&&
f_3(\infty)=0\, , \qquad   f(\infty) = 0
\label{fbc}
\eeqn
for the gauge fields, while the boundary conditions for  the
squark fields are
\beqn
\phi_1 (\infty)=\sqrt{\xi}\,,\qquad   \phi_2 (\infty)=\sqrt{\xi}\,,
\qquad \phi_1 (0)=0\, .
\label{phibc}
\eeqn
Note that since the field $ \phi_2 $ does not wind, it need not vanish
at the
origin, and it does not. Numerical solutions of the
Bogomolny equations (\ref{foest}) for the (0,1) and (1,0) strings were
found in
Ref.~\cite{Auzzi:2003fs}, see e.g. Figs. 1 and 2 in this paper.

\subsection{Non-Abelian moduli}
\label{nonabelianmoduli}

\begin{figure}[h]
\epsfxsize=5cm
\centerline{\epsfbox{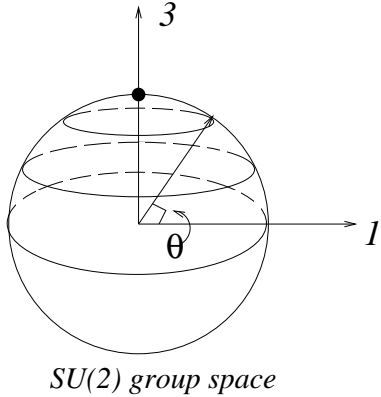}}
\caption{Unwinding the (1,-1)-string.}
\label{vstone}
\end{figure}

Now let us assume  $\Delta m =0$ and demonstrate the occurrence
of a more general solution \cite{Auzzi:2003fs} which contains
non-Abelian moduli. The adjoint scalar VEV does not break
the gauge SU(2) if  $\Delta m =0$.
The relevant  homotopy group in this case is the fundamental group
\beq
\label{pi1suu}
\pi_1\left( \frac{{\rm SU}(2)\times {\rm  U}(1)}{ Z_2} \right) =  Z\, .
\eeq
This means that the $(n,k)$-string lattice reduces to a tower
labeled by a single integer $(n+k)$. For instance, the
$(1,-1)$ string becomes
classically unstable (no barrier).  On the SU(2) group manifold
it corresponds to a winding along the equator on the sphere $S_3$
(Fig.~\ref{vstone}). Clearly this winding can be shrunk to zero
by contracting the loop towards north or south poles \cite{SYmeta}.
On the other hand, the elementary $(1,0)$ and $(0,1)$ strings
cannot be shrunk, as was   explained above.
They correspond to a half-circle winding along the equator. The
$(1,0)$ and $(0,1)$ strings
form a doublet of the residual global SU(2)$_{C+F}$.

A remarkable feature of the $(1,0)$ and $(0,1)$ strings is the occurrence
of non-Abelian moduli which are absent in the ANO strings.
Indeed, while the vacuum field $\Phi_{\rm vac} =\sqrt\xi \, I$
(here $I$ is $2\times2$ unit matrix)  is invariant
under the global SU(2)$_{C+F}$,
\beq
\Phi \to U_L \Phi U_R\,,\qquad U_R= U_L^\dagger\,,
\eeq
the string configuration (\ref{sol}) is not.\footnote{Below we will drop
the subscripts $R$ and $L$.}
Therefore, if there is a solution of the form
(\ref{sol}), there is in fact
a family of solutions obtained from (\ref{sol}) by
the combined global gauge-flavor rotation. Say, for quark fields
it reads
\beq
\Phi (x) \to e^{i\beta}e^{i\vec \omega\, \vec\tau /2} \,
\Phi (x) \, e^{- i\beta}e^{-i\vec \omega\,
\vec\tau /2} \,.
\label{bmod}
\eeq
 The U(1) factor $e^{i\beta}$ does not act on the string solution
(\ref{sol}). It is not to be counted. Thus, what remains is SU(2).
In fact, it is the coset SU(2)/U(1), as is rather obvious from
Eq.~(\ref{sol}):  rotations around the third axis in the SU(2) space
leave the solution with the asymptotics  (\ref{sol})  intact.

Thus, the introduction of the ``moduli matrix" $U$ allows one to get
a generic solution of the non-Abelian string Bogomolny equation
with the following asymptotics
at $|x|\to\infty$:
\beq
\Phi (x) = \sqrt\xi \exp\left( i \frac{\alpha (x)}{2} \right)\,
\exp \left( i\alpha (x) \frac{\vec n\vec\tau}{2} \right)\,,
\label{newwip}
\eeq
where $\vec n$ is a moduli vector
defined as
\beq
n^a\tau^a=U \tau^3 U^{-1},\;\;a=1,2,3.
\label{n}
\eeq
It is  subject to the condition
\beq
{\vec n}^{\,2} = 1\,.
\label{ensq}
\eeq
At $n=\{0,0,\pm 1\}$ we get the field configurations quoted
in Eq.~(\ref{newwi}). Every given matrix $U$ defines the moduli vector
$\vec n$
unambiguously. The inverse is not true. If we consider the left-hand side
of Eq.~(\ref{n})
as given, then the solution for $U$ is obviously ambiguous, since for any
solution
$U$ one can construct two ``gauge orbits" of solutions, namely,
\beqn
U
&\to&
 U\, \exp\left( i\alpha \tau_3\right)\,,
\nonumber\\[3mm]
U
&\to&
\exp\left( i\beta \vec n\vec  \tau \right)\,  U\,.
\label{gaor}
\eeqn
We will use this freedom in what follows.

At finite $|x|$ the non-Abelian string centered at the origin
can be written as \cite{Auzzi:2003fs}
\begin{eqnarray}
\Phi (x)
&=&
U \left(
\begin{array}{cc}
 e^{  i \alpha}   \phi_1(r) & 0  \\[2mm]
0 &  \phi_2(r)
\end{array}
\right)U^{-1}
\nonumber \\[5mm]
 & =&
 e^{\frac{i}{2}\alpha  (1+n^a\tau^a)} \,
 U\left(
\begin{array}{cc}
 \phi_1(r) & 0  \\[2mm]
 0 &  \phi_2(r)
 \end{array}
 \right)U^{-1}\,,
\nonumber \\[5mm]
A^a_{i}(x)  &=& -\,n^a \,\varepsilon_{ij}\,\frac{x_j}{r^2}\,
[1-f_3(r)] \, ,
\nonumber \\[5mm]
\label{rna}
A_{i}(x) &=& -\varepsilon_{ij}\,\frac{x_j}{r^2}\,
[1-f(r)] \,,
\end{eqnarray}
where the profile functions are the solutions to Eq.~(\ref{foest}).
Note that
\beq
U \left(
\begin{array}{cc}
   \phi_1  & 0  \\[1mm]
0 &  \phi_2
\end{array}
\right)U^{-1} =\frac{\phi_1+\phi_2}{2} +n^a\,\tau^a\,
\frac{\phi_1-\phi_2}{2}\,.
\label{uphiu}
\eeq
Now it is   particularly clear that  this solution smoothly
interpolates between  the $(1,0)$ and
$(0,1)$ strings:   if $n=(0,0,1)$ the
first-flavor $r$ squark  winds at infinity while for $n=(0,0,-1)$ it is the
second-flavor  $b$ squark.

Since  the SU(2)$_{C+F}$  symmetry is not broken by the squark vacuum
expectation values,  it is physical and has nothing to do with
the gauge rotations eaten  by the Higgs mechanism. The orientational moduli
$\vec n$ are {\em not} gauge artifacts.
To see this  we can construct {\em gauge invariant} operators which
have explicit  $\vec n$-dependence. Such a construction
 is convenient in order to elucidate  features of our
non-Abelian string solution as well as for other purposes.

\begin{figure}
\epsfxsize=4cm
\centerline{\epsfbox{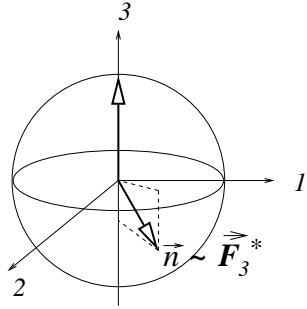}}
\caption{
Bosonic moduli $n^a$ describe the orientation of
the color-magnetic flux of the (0,1) and (1,0) strings in the O(3)
group space, Eq.~(\ref{ginvF}).}
\label{mfrid}
\end{figure}

As an example, let us  define the
``non-Abelian" field strength (to be denoted by bold letters),
\beq
{\bf{F}}_3^{*a} =\frac{1}{\xi}\,{\rm Tr}
\left(\Phi^\dagger F_3^{*b}\frac{\tau^b}{2}\Phi\, \tau^a
\right)\,,
\label{gidefi}
\eeq
where the subscript 3 marks the $z$ axis, the direction of the string
(Fig.~\ref{mmmon}).
From the very definition it is clear that this field
is {\em gauge invariant}.\footnote{In the vacuum,
where the matrix $\Phi$ is that of VEV's,
${\bf{F}}_3^{*a}$ and $ F_3^{*a}$ would coincide.}
Moreover,
Eq.  (\ref{rna}) implies that
\beq
{\bf{F}}_3^{*a} =- n^a\, \frac{(\phi_1^2+\phi_2^2)}{2\xi}\, \frac1r \,
\frac{df_3}{dr}\,.
\label{ginvF}
\eeq
From this formula we readily infer the physical meaning of the moduli
$\vec n$:
the flux of the {\em color}-magnetic field\,\footnote{Defined in the
gauge-invariant way, see Eq.~(\ref{gidefi}).} in the flux tube
is directed along $\vec n$ (Fig.~\ref{mfrid}).
For strings in  Eq.~(\ref{sol}), see also Eq.~(\ref{newwi}),  the
color-magnetic flux is
directed along the third axis in the O(3) group space, either upward
or downward.
It is just this aspect that allows us to refer to the strings above
as ``non-Abelian."  They are as non-Abelian as it gets at weak coupling.

Although the flux tubes in non-Abelian theories at weak coupling
were studied in numerous papers in
recent years \cite{VS,HV,Su,SS,KB,KoS,MY},
in the previous constructions the flux was always directed in a
fixed group direction (corresponding to a Cartan subalgebra), and no moduli
that would freely govern its orientation in the group space were ever
obtained.

To conclude this  section let us present the non-Abelian string
solution (\ref{rna}) in the singular gauge in which the squark fields
at $r\to \infty$ tend to   fixed VEV's and do not wind (i.e. do not
depend on the polar angle $\alpha$ at
$|x|\to\infty$ in the $x_{1,2}$ plane). In the singular gauge we have
\begin{eqnarray}
\Phi
&=&
U \left(
\begin{array}{cc}
\phi_1(r) & 0  \\[2mm]
0 &  \phi_2(r)
\end{array}\right)U^{-1}\, ,
\nonumber \\[4mm]
A^a_{i}(x)
&=&
n^a \,\varepsilon_{ij}\, \frac{x_j}{r^2}\,
f_3(r)\, ,
\nonumber \\[4mm]
A_{i}(x)
&=&
 \varepsilon_{ij} \, \frac{x_j}{r^2}\,
f(r)\, .
\label{sna}
\end{eqnarray}
In this gauge the spatial components of $A_\mu$ fall fast at large distances.
If  the chromo-magnetic flux is defined as a circulation of $A_i$ over
a circle encompassing the string axis, the flux will be saturated by the
integral
coming from the small circle around the (singular) string origin.
We will use this  singular-gauge form of the string solution later.

\subsection{Supersymmetry}
\label{subsupersymmetry}

So far the \ntwo SUSY nature of the model at hand was behind the scene.
Now it is in order to discuss this aspect.
The central charge relevant to flux tubes is the so-called (1/2,1/2)
central charge \cite{GS};
in \ntwo theory it can be written as follows \cite{VY}:
\beq
\{Q_\alpha^f\, \bar Q_{\dot\beta\, g}\} =
2\delta^f_g (\sigma_{\mu})_{\alpha\dot\beta}P_{\mu}
+4i(\sigma_{\mu})_{\alpha\dot\beta}\, \xi^f_g\, \int d^3 x \frac12
\varepsilon_{0\mu\delta\gamma}F_{\delta\gamma},
\label{1212cc}
\eeq
where $P_{\mu}$ is energy-momentum operator,  while\,\footnote{The
factor $i$ in the second term on the right-hand side is due
to our ``Euclidean" notation, see footnote to Eq.~(\ref{redqed}).
Note that the definition of $\xi^f_g$ in Ref.~\cite{VY}
differs by a factor of 2.}
\beq
\xi^f_g=(\tau^m/2)^f_g\, \xi^m\,,
\label{vyvector}
\eeq
where $\xi^m$ is an  SU(2)$_R$ triplet
of the Fayet-Iliopoulos parameters. In the model under consideration
only $\xi^1\equiv \xi$ is non-zero, so that
\beq
\xi^f_g = \frac{\xi}{2}\,(\tau_1)^f_g\,.
\eeq
The central charge is the second term in Eq.~(\ref{1212cc}).
Note that it is only the U(1) field strength tensor that appears
in this central charge. Moreover, it is obvious that
\beq
\int d^3 x \frac12
\varepsilon_{0\mu\delta\gamma}F_{\delta\gamma} = L\, n_\mu \, \int dx\,dy\,
 F_3^{*}=
Ln_\mu \times {\rm (Magnetic \,\,\, Flux)}\,=2\pi\, L\, n_\mu\, ,
\label{ccdop}
\eeq
where $L$ is the string length and $n_\mu$ is the unit vector pointing
in the
direction of the string axis (which coincides with the $z$
axis in our convention, see Fig.~\ref{mmmon}.)
The strings in question are 1/2 BPS-saturated so that
(i) the combination of equations
(\ref{1212cc}) and (\ref{ccdop}) implies that the tensions
(\ref{pmst}) are exact, and (ii)
 a ``macroscopic" world-sheet theory must have
four supercharges, i.e. must be \ntwo in terms of two-dimensional
world-sheet classification. This is because the original ``microscopic"
theory has eight supercharges.
Note that the Fayet-Iliopoulos parameter is not renormalized by
quantum corrections.

The flux tube is characterized by two translational moduli,
$x_0$ and $y_0$, the positions of the string center in the
perpendicular plane.
The four supercharges that act non-trivially supplement
$x_0$ and $y_0$ by four fermionic superpartners $\eta^i$
($i=1,2,3,4$), supertranslational moduli.
As usual, the (super)translational moduli split (i.e. completely decouple)
from those
describing internal dynamics. The dimension of representation is four,
but this is a trivial multiplicity which we do not count
when we speak of  distinct elementary strings.
In fact, a ``distinct" string
means that we deal with  distinct dimension-four supermultiplets.

We will return to  ${\cal N}=2$ superalgebra,
in the anticommutator $\{Q_\alpha^f\,  Q_{\ \beta}^g\}$
 relevant to the  issue of string junctions
(confined monopoles), in Sect.~\ref{quantumlimit}.

\section{Macroscopic theory}
\label{macroscopictheory}
\setcounter{equation}{0}

In this section we will derive an effective low-energy
theory for orientational collective coordinates on the string world sheet.
As was already mentioned, this macroscopic theory
is  a two-dimensional supersymmetric $CP^1$ model. This model is discussed
in great detail in the review paper \cite{NSVZ} from which we borrow
many definitions, notation and normalizations.
At first we will review  derivation  \cite{Auzzi:2003fs} of  the effective
theory for the bosonic moduli $n^a$. In Sects.~\ref{kineticterm},
\ref{11dimensional} the mass difference is set to zero, $\Delta m =0$.
Switching on $\Delta m $ will lead us to the $CP^1$ model with a
``twisted mass." Then we will work out the fermionic part.

\subsection{Deriving the kinetic term; a basic normalizing integral}
\label{kineticterm}

Assume  that the orientational collective coordinates $n^a$
are slow varying functions of the string world-sheet coordinates
$x_k$, $k=0,3$. Then the moduli $n^a$ become fields of a
(1+1)-dimensional sigma model on the world sheet. Since
the vector $n^a$ parametrizes the string zero modes,
there is no potential term in this sigma model.  We begin with
the kinetic term \cite{Auzzi:2003fs}.

To obtain the   kinetic term  we substitute our solution, which depends
on the moduli $\vec n$,
in the action (\ref{redqed}) assuming  that
the fields acquire a dependence on the coordinates $x_k$ via $n^a(x_k)$.
Technically it is convenient to work with the solution
(\ref{sna}) in the singular gauge. In
doing so we immediately observe that we {\em must} modify the solution.

Indeed, Eq.~(\ref{uphiu}) or (\ref{sna})
were obtained as an SU(2)$_{C+F}$ rotation of the ``basic"
$(1,0)$-string (\ref{sol}).
Now we make this transformation local (i.e. depending on $x_k$).
Because of this,
the 0 and $z$ components of the gauge potential  no longer vanish. They
must be added to  the {\em ansatz}. This situation is quite familiar (e.g.
\cite{SYmeta}) since one routinely encounters it in the soliton studies.

We suggest an obvious {\em ansatz} for these components
(to be checked {\em a posteriori}),
\beq
A_k=-i\,  \left( \pt_k U\right) \,U^{-1}\,\rho (r)\, , \qquad k=0, 3\,,
\label{An}
\eeq
where a new profile function $\rho (r)$ is introduced.
To be more precise, we must express the right-hand side in terms of our
moduli fields $n^a$. As was mentioned after Eq.~(\ref{ensq}),
the parametrization of the matrix $U$ is ambiguous.
Correspondingly, if we introduce
\beq
\alpha_k \equiv -i\,  \left( \pt_k U\right) \,U^{-1}\,,
\qquad \alpha_k \equiv \alpha_k^a\,\left( \frac{\tau^a}{2}\right) \,,
\label{alphak}
\eeq
then $\alpha_k^a$ is defined modulo two gauge transformations following
from
Eq.~(\ref{gaor}). Equation (\ref{n}) implies that
\beq
\alpha_k^a - n^a \, \left( n^b \, \alpha_k^b\right) =
- \varepsilon^{abc}\, n^b\, \partial_k n^c \,,
 \label{nimlic}
\eeq
and we can impose the condition $ n^b \, \alpha_k^b$=0. Then
\beq
\alpha_k^a
=
-   \varepsilon^{abc}\, n^b\, \partial_k n^c \,,
\qquad
-i\,  \left( \pt_k U\right) \,U^{-1} = -\frac{1}{2}\, \tau^a\,
\varepsilon^{abc}\, n^b\, \partial_k n^c \,.
\label{endan}
\eeq

The function $\rho (r)$ in Eq.~(\ref{An}) is
determined  through a minimization procedure (cf. \cite{SYmeta})
which generates $\rho$'s own equation of motion. Our task is to derive it.
But at first we note that
 $\rho (r)$ vanishes at infinity,
\beq
\rho (\infty)=0\,.
\label{bcfinfty}
\eeq
The boundary condition at $r=0$ will be determined shortly.

The kinetic term for $n^a$ comes from the gauge and quark kinetic terms
in Eq.~(\ref{redqed}). Using (\ref{sna}) and (\ref{An}) to calculate the
SU(2)  gauge field strength we find
\beq
\label{Fni}
F_{ki}=\frac12\,\left( \pt_k n^a\right) \tau^a\,
\varepsilon_{ij}\,\frac{x_j}{r^2}\,
f_3
\left[1-\rho (r)\right]+
i\left(\pt_k U\right) \,U^{-1}\,\frac{x_i}{r}\,\, \frac{d\,\rho (r)}{dr}\, .
\eeq
 We see that in order to have a finite contribution  from
Tr$F_{ki}^2$ in the action we have to impose the constraint
\beq
\rho (0)=1\,.
\label{bcfzero}
\eeq
Substituting the field strength (\ref{Fni}) in the action
(\ref{redqed}) and including, in addition, the  kinetic term of quarks, we
arrive at
\beq
S^{(1+1)}= \frac{ \beta}{2}\,   \int d t\, dz \,
\left(\pt_k\,  n^a\right)^2\,,
\label{o3}
\eeq
where the coupling constant $\beta$ is given by a normalizing integral
\beqn
\beta
&= &
\frac{2\pi}{g_2^2}\,  \int_0^{\infty}
rdr\left\{\left(\frac{d}{dr}\rho (r)\right)^2
+\frac{1}{r^2}\, f_3^2\,\left(1-\rho \right)^2
\right.
\nonumber\\[4mm]
&+& \left.  g_2^2\left[\frac{ \rho^2}{2}\left(\phi_1^2+\phi_2^2\right)
+\left(1-\rho \right)\left(\phi_1-\phi_2\right)^2\right]\right\}\, .
\label{beta}
\eeqn
We will have to deal with the integral on the right-hand side
more than once in what follows. There are various arguments allowing one
to find it analytically; however, the first calculation was carried
out numerically by  R. Auzzi to whom we are deeply grateful.

The functional (\ref{beta}) must be minimized with respect to
$\rho$ with the boundary conditions given by
(\ref{bcfinfty}), (\ref{bcfzero}). Varying (\ref{beta}) with respect to
$\rho$
one readily obtains the second-order equation which
the function $\rho$ must satisfy,
\beq
-\frac{d^2}{dr^2}\, \rho -\frac1r\, \frac{d}{dr}\, \rho
-\frac{1}{r^2}\, f_3^2 \left(1-\rho\right)
+
\frac{g^2_2}{2}\left(\phi_1^2+\phi^2_2\right)
\rho
-\frac{g_2^2}{2}\left(\phi_1-\phi_2\right)^2=0\, .
\label{rhoeq}
\eeq
The second-order equation occurs  because,
once  we allow the  dependence of $n^a$ on the world-sheet coordinates
$t,z$, the string is no longer BPS saturated.
After some algebra and extensive use of the first-order equations
(\ref{foest})
one can show that the solution of (\ref{rhoeq})  is given by
\beq
\rho=1-\frac{\phi_1}{\phi_2}\, .
\label{rhosol}
\eeq
This solution  satisfies the boundary conditions (\ref{bcfinfty})
and  (\ref{bcfzero}), as it should.

Substituting this solution back into the expression for the
sigma model coupling constant (\ref{beta}) one can check that
the integral  in (\ref{beta}) reduces to a total derivative and given
by the flux of the string  determined by $f_3(0)=1$.
 Namely,\footnote{The numerical
result of  R. Auzzi mentioned above was $I=1.00008$.}
\beqn
I
&\equiv&
  \int_0^{\infty}
rdr\left\{\left(\frac{d}{dr}\rho (r)\right)^2
+\frac{1}{r^2}\, f_3^2\,\left(1-\rho \right)^2
\right.
\nonumber\\[4mm]
&+& \left.  g_2^2\left[\frac{ \rho^2}{2}\left(\phi_1^2+\phi_2^2\right)
+\left(1-\rho \right)\left(\phi_1-\phi_2\right)^2\right]\right\}
\nonumber\\[4mm]
&=&\int_{0}^{\infty}  dr\left(-\frac{d}{dr}f_3\right) = 1,
\label{norin}
\eeqn
where we use the first order equations (\ref{foest}) for the profile
functions of the string.
Returning to the $CP^1$ model (\ref{o3})
we conclude that the
sigma model coupling $\beta$ does not depend on the ratio of U(1) and
SU(2) coupling constants and is given by
\beq
\beta= \frac{2\pi}{g_2^2}\,.
\label{betagamma}
\eeq
The two-dimensional coupling constant is determined by the
four-dimensional non-Abelian coupling.
As we will see later,   this fact is very important for our
interpretation of  confined monopoles  as  sigma-model kinks.

In summary,  the effective world-sheet theory describing dynamics
of  the string orientational zero modes is the celebrated
$O(3)$ sigma model (which is the same as $CP^1$).  The symmetry
of this model reflects the presence of the  global SU(2)$_{C+F}$
symmetry in the microscopic theory.
The coupling constant of this sigma model is
determined by  minimization of the action  (\ref{beta}) for the
profile function $\rho$. The minimal value of $I$ is unity.
Clearly, Eq.~(\ref{o3}) describes the low-energy limit.
In principle, the zero-mode interaction  has higher derivative
corrections which run in powers of
\beq
\left( g_2\, \sqrt{\xi}\right)^{-1} {\pt_n} \,,
\label{hd}
\eeq
where $g_2\sqrt{\xi}$ gives the order of magnitude of
masses of the
gauge/quark multiplets in our microscopic SU(2)$\times$U(1)  theory.
The sigma model (\ref{o3}) is adequate at  scales
 below
$ g_2\sqrt{\xi}$ where higher-derivative corrections are negligibly small.

The very same scale, $\left( g_2\, \sqrt{\xi}\right)^{-1}$,
determines   the   thickness of the strings we deal with.
In other words,
the effective sigma model  (\ref{o3}) is applicable
at scales below the inverse string thickness
which, thus,  plays the role of an  ultraviolet (UV) cutoff for the
model (\ref{o3}).

\subsection{(1+1)-dimensional $CP^{1}$ model}
\label{11dimensional}

Let us discuss the theory on the string world sheet as it emerges
after factoring out (super)translational moduli.
As was mentioned, the solution of the string BPS condition
is in fact a two-parametric family of solutions parametrized by
$\vec n$ with the constraint  ${\vec n}^2 =1$. The target
space of the bosonic moduli is SU(2)/U(1), the same as $CP^1$.

Since it is also endowed with four supercharges,
the world-sheet theory must be  \ntwo
two-dimensional $CP^1$ sigma model. In Sect.~\ref{fermionzero}  we will
explicitly construct four fermion zero modes in the microscopic theory
(not counting supertranslational) which  match two bosonic zero modes
associated with the color-magnetic flux rotation in the O(3)
group space, Fig.~\ref{mfrid}. This will essentially conclude the proof
that the world-sheet theory is the $CP^1$ sigma model. Let us briefly
review the properties of this model.

The \ntwo two-dimensional $CP^1$ model has the following action
(e.g. \cite{NSVZ}, Sect. 6):
\beq
S_{CP(1)}=\beta \int d^2 x \left\{\frac12 \left(\pt_k
n^a\right)^2 + \frac{i}{2} \, \bar{\chi}^a \gamma_k\pt_k \chi^a+
\frac18\, (\bar{\chi}\chi)^2  \right\}  \, ,
\label{o3n2}
\eeq
where the $\gamma$ matrices in (1+1) dimensions are
defined as\,\footnote{
This is a  ``Euclidean" notation, see footnote to Eq.~(\ref{redqed}).}
\beq
\gamma_0=\tau_1,\;\;\; \gamma_3=-\tau_2\, ,
\label{gammamat}
\eeq
while $\chi^a$ is a real two component Majorana fermion field
($\bar{\chi}=\chi \gamma_0$). It is subject to the
constraint
\beq
\chi^a n^a=0\,;
\label{chiconstr}
\eeq
therefore,  in fact, we have  four real fermion components, as expected.
Note that $\beta$ is related to the conventional
coupling constant of the $CP^1$ model as
$\beta = 1/g^2_{CP(1)}$, and  the loop expansion runs in powers of
$g^2_{CP(1)}/\pi$. Given Eq.~(\ref{betagamma}) we see that
that the loop expansion parameter is $g_2^2/(2\pi^2)$ which coincides
with the loop expansion parameter in the microscopic theory.

In the holomorphic representation  upon the stereographic
projection the Lagrangian of the $CP^1$  model (\ref{o3n2})
becomes
\beqn
&& {\cal L}_{ CP(1)}= G\, \left\{
\partial_k\bar{w}\, \partial_k w
+ \frac{i}{2}\left(\bar\Psi_L\stackrel{\leftrightarrow}
{\partial_R}\Psi_L + \bar\Psi_R\stackrel{\leftrightarrow}{\partial_L}\Psi_R
\right)
\right.
\nonumber\\[3mm]
&&-\frac{i}{\zeta}\, \left. \left[\bar\Psi_L \Psi_L
\left(\bar{w}\stackrel{\leftrightarrow}{\partial_R} w
\right)+ \bar\Psi_R \Psi_R
\left(\bar{w}\stackrel{\leftrightarrow}{\partial_L} w
\right)
\right]
-
\frac{2}{\zeta^2}\bar\Psi_L\Psi_L \bar\Psi_R\Psi_R
\right\}
\nonumber\\[3mm]
&&+
\frac{i \theta}{2\pi}\,\frac{1}{\zeta^2}\,
\varepsilon^{mk} \partial_m\bar{w}\,
\partial_k w\,,
\label{13one}
\eeqn
where  $G$ is the metric on the target space,
\beq
G  \equiv 2\beta\,\frac{1}{\left( 1+w\bar{w}\right)^2}\, ,
\label{13two}
\eeq
and
\beq
\zeta \equiv 1+w\bar{w} \,.
\label{zetad}
\eeq
(It is useful to note that
the Ricci tensor $R = 2\, \zeta^{-2}$.)
For completeness we also included the vacuum-angle term, see the
last term  in Eq.~(\ref{13one}).
Furthermore, the fermion field is a two-component {\em Dirac} spinor
\beq
\Psi = \left(
\begin{array}{cc}
\Psi_R \\
\Psi_L
\end{array}
\right) \,.
\label{13four}
\eeq
Finally, the bars over $w$ and $\Psi_{L,R}$ denote
Hermitean conjugation.

The sigma model (\ref{o3n2}) or (\ref{13one})
is asymptotically free \cite{siaf}; at large distances (low energies)
it gets into the strong coupling regime. The corresponding
Gell-Mann--Low function is one-loop, and the running
coupling constant  as a function of
the energy scale $E$ is given by
\beq
4\pi \beta = 2\ln {\left(\frac{E}{\Lambda_{2D}}\right)}
\label{sigmacoup}
\eeq
where $\Lambda_{2D}$ is the dynamical scale of the sigma model;
a related definition of this scale is given below in
Eq.~(\ref{fercond}). As was mentioned previously,
the ultraviolet cut-off of the sigma model at hand
is determined by  $g_2\sqrt{\xi}$. At this UV cut-off scale
Eq.~(\ref{betagamma}) holds.
Hence,
\beq
\Lambda^2_{2D} = \xi e^{-\frac{8\pi^2}{g_2^2}} =\Lambda^2\, ,
\label{lambdasig}
\eeq
where we take into account Eq.~(\ref{alter}) for the dynamical scale
$\Lambda$ of the SU(2) factor of the microscopic theory.
Note that in the microscopic theory {\em per se}, because of the VEV's of
the squark fields, the coupling constant is frozen at
$g_2\sqrt{\xi}$; there are no logarithms below this scale.
The logarithms of the macroscopic theory take over.
Moreover, the dynamical scales of the microscopic and microscopic
theories turn out to be the same!
We will explain the reason why the dynamical
scale of the (1+1)-dimensional effective theory
on the string world sheet   equals that
of the SU(2) factor of the (3+1)-dimensional gauge theory
later, in Sect.~\ref{dopo6}.

The superalgebra induced by four supercharges of the world-sheet theory
is as follows:
\beqn
\{\bar Q_L  Q_L\} &=&  \left(
H+ P\right)
\,,\quad \{\bar Q_R  Q_R\} = \left(  H-P\right)\,;
\label{13eight}
\\[3mm]
\{\bar  Q_R Q_L \} &=&
\frac{ 1}{ \pi}  \int dz\, \partial_z \left(\zeta^{-2}\,
\bar\Psi_R  \Psi_L \right) \,,
\label{13twelve}
\\[3mm]
\{\bar  Q_L Q_R \}
&=& \frac{1}{\pi}  \int dz\, \partial_z \left( \zeta^{-2}\,
\bar\Psi_L  \Psi_R \right) \,.
\label{13thirteen}
\eeqn
with all other anticommutators vanishing.
Here $(H,P)$ is the energy-momentum operator.
Equations (\ref{13twelve}) and (\ref{13thirteen})
present a quantum anomaly derived in Ref.~\cite{Losev}. These
anticommutators vanish at the classical level.
The above anomaly is similar
(and, in fact, related) to that in \none supersymmetric
gluodynamics \cite{dwano}.
Its  occurrence  is crucial for self-consistency of
matching of the underlying microscopic theory with the macroscopic
description provided by Eq.~(\ref{13one}). The fact
that  the anomaly does take place
can be viewed as a test that we are on the right track.

As well-known, two-dimensional  $CP^1$ model possesses
two vacua labeled by the bifermion order parameter,
\beq
\langle \zeta^{-2}\bar\Psi_R  \Psi_L \rangle =\pm \Lambda_{\, CP(1)} \,
e^{i\theta/2}\,.
\label{fercond}
\eeq
The  distinct vacua of the world-sheet effective theory, in the language
of the microscopic theory, describe two distinct strings.
The physical meaning of this distinction will be revealed shortly.
The dynamical scale $\Lambda_{2D}$ defined in Eq.~(\ref{sigmacoup})
is of the order of $\Lambda_{\, CP(1)}$. More exactly,
$\Lambda_{2D} = e\, \Lambda_{\, CP(1)}$.

We will interrupt here our discussion of the superalgebra in the macroscopic
theory, with the intention to return to it later, in
Sect.~\ref{quantumlimit}.

\subsection{Unequal quark  mass terms; $CP^1$ with the twisted mass}
\label{unequalquark}

The fact that we have two distinct vacua in the world-sheet theory
--- two distinct strings ---
is not quite intuitive in the above consideration.
This is   understandable. At the classical level
the \ntwo two-dimensional sigma model has a continuous vacuum
manifold $S_2$. This is in one-to-one correspondence with
continuously many strings parametrized by $\vec n$.
The continuous degeneracy is lifted only upon inclusion of quantum effects
that occur (in the sigma model) at strong coupling. Gone with this lifting
is the moduli nature of the  fields $n^a$. They become massive.
This is difficult to grasp.

To facilitate contact between the microscopic and macroscopic theories,
it is instructive to start from a deformed  microscopic theory
so that the string moduli are lifted already at the classical level.
Then  the origin of the two-fold degeneracy of the
non-Abelian strings become transparent. This will help us
understand, in an intuitive manner, other features
listed above. After this understanding is
achieved, nothing prevents us from returning
to our case of strings with non-Abelian moduli at the classical level,
by smoothly suppressing the moduli-breaking deformation. The two-fold
degeneracy will remain intact as it follows from
the Witten index \cite{WitIndex}.

\begin{figure}[h]
\epsfxsize=9.5cm
\centerline{\epsfbox{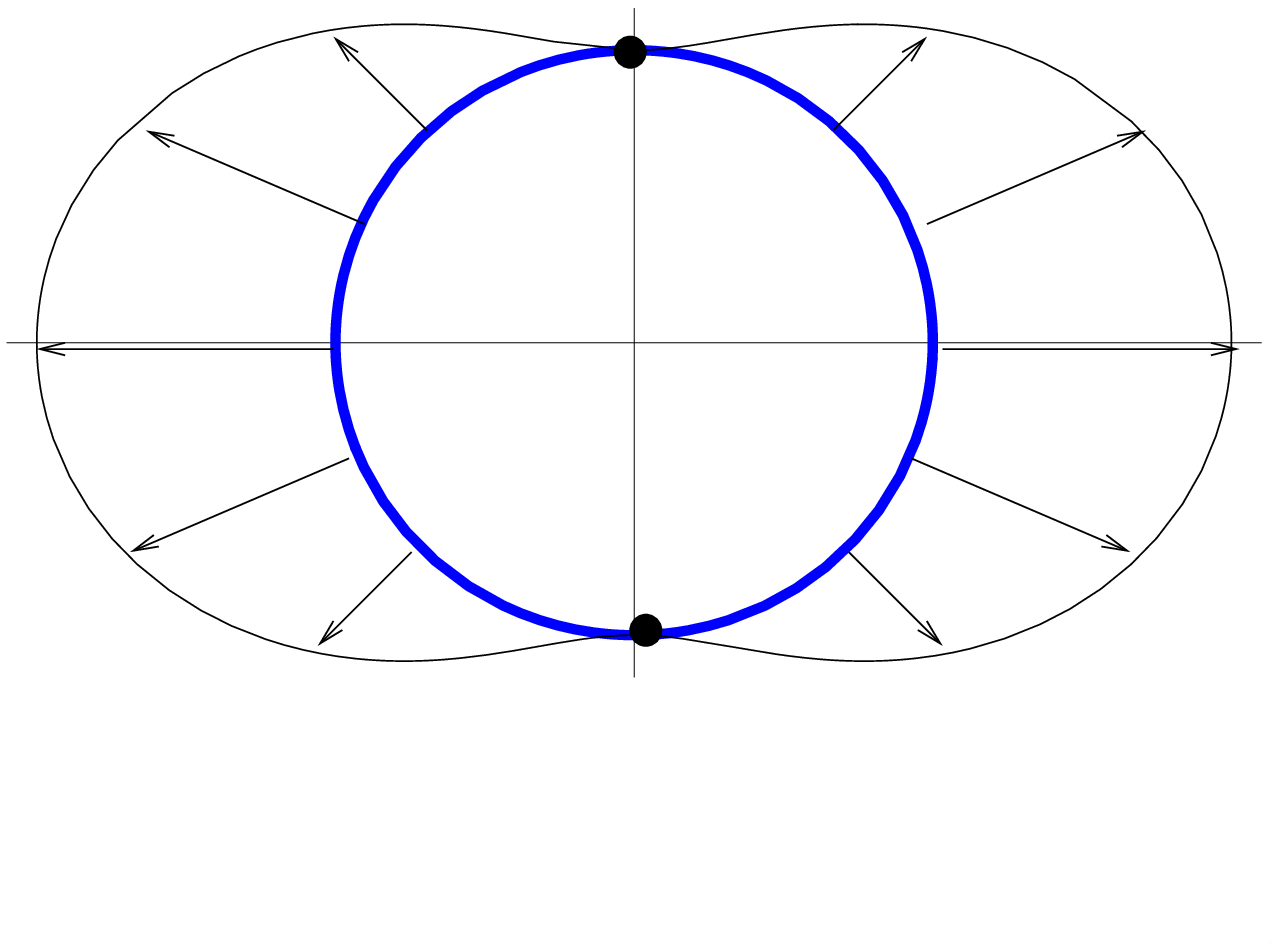}}
 \caption{Meridian slice of the target space sphere (thick solid line).
Arrows present the scalar potential (\ref{13twentythree}),
their length being the strength of the potential.
Two vacua of the model are denoted by closed circles.}
\label{f13one}
\end{figure}

Thus, let us drop the assumption $m_1=m_2$ and introduce a
small mass difference. We will still assume that
\beq
\mu \equiv m_1-m_2 > 0\, .
\eeq
At $m_1\neq m_2$ the flavor (global)  SU(2) symmetry
of the microscopic theory is explicitly broken
down to U(1) (corresponding to rotations around the third axis in the
O(3) group space).  Correspondingly,
the moduli of the non-Abelian string are lifted, since the vector $\vec n$
gets fixed in the position pointing in the third direction,
$\vec n=\{0,0,\pm 1\}$. These are the (1,0) and
(0,1)  strings, respectively, see  Sect.~\ref{elementarystrings}.
If $\mu \ll \sqrt{\xi}$ the set of parameters  $n^a$ becomes {\em quasi}moduli, as is clear from Fig.~\ref{f13one}.

Now, our aim is to derive the effective two-dimensional theory
on the string world sheet for the case of unequal quark masses,
when SU(2)$\times$U(1) gauge theory is broken down to U(1)$\times$U(1),
assuming that $\mu$ is small.  As was discussed in
Sect.~\ref{elementarystrings},
the {\em bona fide} solutions of first-order equations (\ref{foest}) (and,
hence, the equations of motion) with the minimal windings are
the (1,0) and (0,1) strings. The
solution for the (1,0) string is given by (\ref{sol}) while the solution
for the (0,1) string can be obtained from the one in Eq.~(\ref{sol}) by the
replacement
$$
f_3\to -f_3
$$
 and
$$
e^{i\alpha}\phi_1\leftrightarrow \phi_2\,.
$$
However, at  small $\mu$ we can still introduce the orientational
quasi-moduli $n^a$.  In terms of the effective two-dimensional theory on
the string world sheet $\mu\neq 0$ leads to a shallow potential
for the quasi-moduli  $n^a$. The two minima of the
potential  at $n=\{0,0,\pm 1\}$ correspond to two {\em bona fide}
solutions for the (1,0) and (0,1) strings.

Let us derive this potential. To this end  we start from the
expression for the non-Abelian string
in the singular gauge (\ref{sna}) parametrized by moduli $n^a$
and substitute it in the action (\ref{redqed}). The only
modification that we actually have to make is to supplement our
{\it ansatz} (\ref{sna}) by that
for the adjoint scalar field $a^a$;
the neutral scalar field $a$ will stay fixed at its vacuum expectation
value $a=-\sqrt{2}m$.

At large $r$ the field $a^a$ tends to its VEV directed along
the third axis in the
color space and is given by Eq.~(\ref{avev}). At the same time, at $r=0$
it must be directed along the vector $n^a$.  The reason for this behavior
is easy to understand. The kinetic term for $a^a$  in Eq.~(\ref{redqed})
contains the commutator term of the adjoint scalar and the gauge potential.
The gauge potential is singular at the origin, as is seen
from  Eq.~(\ref{sna}). This
implies that $a^a$ must be directed along $n^a$ at $r=0$. Otherwise,
the string tension would become divergent.
The following {\it ansatz} for $a^a$ ensures this behavior:
\beq
a^a=-\frac{\mu}{\sqrt{2}}\, \left[\delta^{a3}\, b +n^a\,  n^3\,
(1-b)\right]\, .
\label{aa}
\eeq
Here we introduced a new profile function $b(r)$ which, as usual, will be
determined from a minimization procedure.
Note that at $n^a=(0,0,\pm 1)$ the field $a^a$ is given by its VEV,
as expected. The boundary conditions for  the function $b(r)$ are
\beq
b(\infty)=1\, ,\qquad  b(0)=0\, .
\label{bbc}
\eeq
Substituting Eq.~(\ref{aa}) in conjunction with (\ref{sna}) in the action
(\ref{redqed})
we get the potential
\beq
V_{CP(1)}=\gamma\int d^2 x\, \frac{\mu^2 }{2} \left(1-n_3^2\right)\, ,
\label{sigpot}
\eeq
where $\gamma$ is presented by  the integral
\beqn
\gamma
&=&
 \frac{2\pi}{g_2^2} \, \int_0^{\infty}
r\, dr\, \left\{\left(\frac{d}{dr}b (r)\right)^2
+\frac{1}{r^2}\, f_3^2\,  b^2+\right.
\nonumber\\[4mm]
&+&
\left.
g_2^2\left[\frac{1}{2}\,  (1-b)^2\, \left(\phi_1^2+\phi_2^2\right)
+b\, \left(\phi_1-\phi_2\right)\right]\right\}\, .
\label{gamma}
\eeqn
Here two first terms in the integrand come from the kinetic term
of  the adjoint scalar field $a^a$  while the term in the square
brackets comes from the  last term in the action (\ref{redqed}).

Minimization with respect to $b(r)$, with the constraint (\ref{bbc}),
yields
\beq
b(r)=1-\rho (r)=\frac{\phi_1}{\phi_2} (r)\,,
\label{bobr}
\eeq
cf. Eqs.~(\ref{beta}),  (\ref{rhosol}). Thus, $\gamma =
I \times 2\pi/(g_2^2)= 2\pi/(g_2^2)$.
We see that the normalization integrals are the same for both, the kinetic
and the potential terms in the world-sheet sigma model, $\gamma=\beta$.
As a result we arrive at  the following effective theory on the string world
sheet:
\beq
\label{o3mass}
S_{CP(1),\,\mu}=\beta \int d^2 x \left\{\frac12 \left(\pt_k
n^a\right)^2 +\frac{\mu^2}{2}\,\left( 1-n_3^2\right)\right\}\, .
\eeq
This is the only functional form that allows \ntwo completion.\footnote{
Note, that although the global SU(2)$_{C+F}$ is broken by $\Delta m$,
the extended ${\cal N}=2$  supersymmetry is not. }

The fact that we obtain this form shows that our {\em ansatz} is
fully adequate. The informative aspect of the procedure
is (i) the confirmation of the {\em ansatz} and (ii)
constructive calculation of the constant
in front of $\left(1-n_3^2\right)$ in terms of the microscopic parameters.
The mass-splitting parameter $\Delta m$ of the microscopic
theory exactly coincides
with the twisted mass $\mu$ of the macroscopic model.

As was already mentioned this sigma model gives an effective
description of our string at low energies, i.e. energies much lower
than the inverse string thickness. Typical momenta in the theory
(\ref{o3mass}) are of the order of $\mu$. Therefore,
for the action (\ref{o3mass})  to be applicable we must
impose the condition
\beq
\left| \Delta m \right| \ll g_2\sqrt{\xi} \, .
\label{mxi}
\eeq
The $CP^1$ model (\ref{o3mass}) has two vacua located at $n^a=(0,0,\pm
1)$, see Fig. \ref{f13one}. Clearly these two vacua correspond
to two elementary strings: (1,0) and (0,1), respectively.

Upon stereographic projection the action (\ref{o3mass}) takes the form
\beq
S_{\rm CP(1),\,\mu }= \int d^2 x \,G\, \left\{
\partial_k\bar{w}\, \partial_k w +\mu^2 \, |w|^2\right\} \,,
\label{o3massw}
\eeq
where $G$ is given in Eq.~(\ref{13two}). We pause here to make
a few remarks
regarding the sigma model with $\mu\neq 0$.

First and foremost, Eq.~(\ref{o3massw}) is the bosonic part
of an \ntwo two-dimensional sigma model \cite{Alvarez}
which is usually referred to as the $CP^1$ model with the twisted mass.
This is a generalization of the massless $CP^1$ model which
preserves four supercharges. As we know, the BPS nature of the
strings under consideration does require the world-sheet
theory to have four supercharges. Crucial for the construction of the
twisted-mass \ntwo model is the fact that
the target space of the $CP^1$ model has
isometries.  One can exploit the isometries to introduce the \ntwo
supersymmetric mass term $\mu$, namely,
\beq
\Delta_\mu {\cal L}_{\rm CP(1)}
= G\left\{
|\mu |^2\, w\bar{w}
+ i\, \frac{1-w\bar{w}}{\zeta}
\left(\mu\, \bar\Psi_L\Psi_R - \bar{\mu} \, \bar\Psi_R\Psi_L
\right)\right\}\,,
\label{13sixteen}
\eeq
to be added to Eq.~(\ref{13one}).
Here $\zeta$ is defined in Eq.~(\ref{zetad}).
Generically speaking, $\mu$ is a complex parameter.
Certainly, one can always
eliminate the phase of $\mu$ by a chiral rotation of the fermion
fields.  Due to the chiral anomaly, this will lead to a shift of the
vacuum angle $\theta$. In what follows
we will keep $\mu$ real and positive and ignore
$\theta$ unless stated to the contrary.
The U(1)-invariant scalar potential term in the holomorphic
representation is
\beq
V_{CP(1),\,\mu}
= \mu^2 \, G\, \bar{w} \, w\,.
\label{13twentythree}
\eeq
If $\mu^2 \gg \Lambda^2_{CP(1)}$, the classical description
is fully applicable.
As we already explained, two minima of $V_{CP(1),\,\mu}$, at $w=0$
and $w=\infty$,
correspond to the (1,0) and (0,1) strings of the four-dimensional theory
(Fig.~\ref{f13one}).

Now, after this lengthy digression,
we can return to the  limit $\Delta m\to 0$. The quasiclassical treatment
of the world-sheet theory applies no longer since the
world-sheet theory gets into the
strong coupling regime, but we  still have two vacua
(Witten's index!).
These two vacua differ from each other by the expectation value
of the chiral bifermion operator (\ref{fercond}), see e.g. \cite{NSVZ}.
At strong coupling ($\mu =0$) the chiral condensate is the
order parameter. The $CP^1$ model has a discrete $Z_4$
symmetry, a remnant of the anomalous U(1) chiral symmetry.
The condensate (\ref{fercond}) breaks it down to $Z_2$;
hence, the two-fold degeneracy.

\begin{figure}[h]
\epsfxsize=10cm
\centerline{\epsfbox{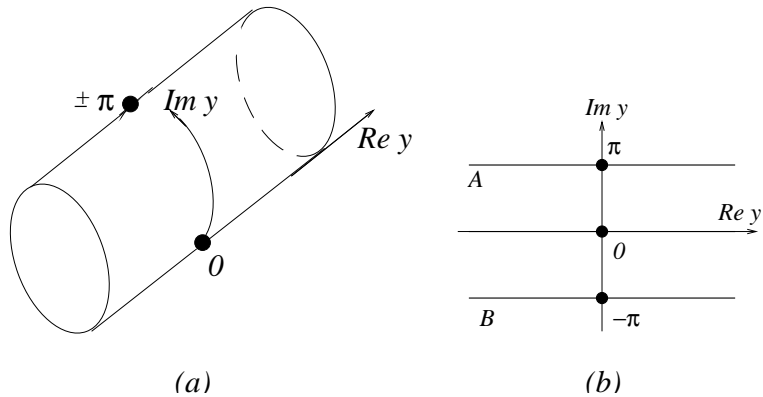}}
\caption{
The target space of the mirror model and the trajectories of the BPS
solutions interpolating between two vacua of the model. The vacua are
denoted by closed circles. The lines $A$ and $B$ in Fig. $b$ must be
identified.}
\label{mirfig}
\end{figure}

The physics of the model becomes more transparent in the mirror
representation  \cite{HoVa}.  In this representation one
describes the $CP^1$ model in terms
of the Coulomb gas of instantons to prove its equivalence to
a sine-Gordon theory. The $CP^1$ model (\ref{o3n2})
is dual to the following \ntwo sine-Gordon model \cite{HoVa}:
\beq
S_{\rm SG}=\int d^2 x  d^2 \theta \,
 d^2 \bar{\theta} \,  {\beta}^{-1} \, \bar{Y}Y +
\left\{ \frac{\Lambda_{CP(1)}}{2\pi}
 \,\int d^2x\,  d^2 \theta \, \cosh{Y} +{\rm h.c.}\right\}\, .
\label{sG}
\eeq
Here the last term is a
dual instanton-induced superpotential.
The scalar potential of this sine-Gordon theory is
\beq
V_{\rm SG}=\frac{\beta }{4\pi^2}\, \Lambda^2_{CP(1)}\, \left|
\sinh {y}\right|^2 \, ,
\label{sGpot}
\eeq
which has two minima, at $y=0$ and $y= \pm i\pi $. The target
space of the mirror model is the cylinder depicted
in Fig.~\ref{mirfig}a. That is why the points $y=   i\pi $ and
$y=   -i\pi $
must be identified; they present one and the same vacuum.
In
Fig.~\ref{mirfig}b this identification means gluing the lines $A$ and
$B$. The straight lines in Fig.~\ref{mirfig}b passing through the
points $y=0$ and $y=\pm i\pi$ are the lines on which the $\cosh Y$
superpotential is real. Since the imaginary part of the superpotential
must vanish on the solutions of the BPS equations starting (ending) on
the vacuum points, the solution trajectories can lie only on the above
lines of  ``reality" of superpotential.

This mirror model explicitly
exhibits  a mass gap of the   order of $\Lambda_{CP(1)}$.  It shows
that there are no Goldstone bosons (corresponding to the absence of the
spontaneous breaking of of the microscopic theory SU(2)$_{C+F}$ on the
string).  This means, in turn, that the string orientation vector $n^a$
has no particular direction, it is smeared all over. The strings we
deal here are genuinely non-Abelian.  Two degenerate strings of the
microscopic theory  corresponding to two degenerate vacua of the theory
(\ref{13one})  are two ``elementary" non-Abelian strings of the $\Delta
m =0$ theory which form a  doublet of SU(2)$_{C+F}$.  They are {\em
not} the $(1,0)$ and $(0,1)$ strings of the quasiclassical
U(1)$\times$U(1) theory since  the vector $n^a$ has no particular
direction.

The mirror model also demonstrates the occurrence of two distinct
kinks interpolating between the two vacua of the model.
(More precisely, since each kink has two fermion zero modes, we should
speak of two distinct dimension-two supermultiplets.) The two
solutions are
$$
y=\pm \left\{\frac{\pi}{2}+{\rm arcsin}[{\rm tanh}\,
m_y z]
\right\}\,,\qquad m_y = \beta \, \Lambda_{CP(1)}/(2\pi )\,.
$$
The
kink doubling is in full accord with the fact that
the Cecotti--Fendley--Intriligator--Vafa (CFIV)  index \cite{CFIV} of
the $CP(1)$ model equals two.

\subsection{Fermion zero modes of the string}
\label{fermionzero}

In this section we use supersymmetry to explicitly construct
the fermion zero modes --- those  not
associated with
supertranslations --- and show that
they reproduce the fermion part of \ntwo two-dimensional
 $CP^1$ model on the string world sheet.
We will set $m_1=m_2$  and  work
with the string solution (\ref{sna}) in the singular gauge.

Our string solution is 1/2 BPS-saturated. This means that four
supercharges, out of eight of the four-dimensional theory, act trivially
on the string solution
(\ref{sna}). The remaining four supercharges generate four fermion zero
modes which are called supertranslational modes because they are
superpartners to  two translational zero
modes. The corresponding four fermionic moduli
are superpartners to the coordinates $x_0$ and
$y_0$ of the string center. The  supertranslational
fermion zero modes were found in Ref.~\cite{VY}.
As a matter of fact, they were found for the U(1) ANO string in
\ntwo  theory but the transition to the model at hand is absolutely
straightforward. We will not dwell on this procedure
here.

Instead, we will focus on four {\em additional} fermion zero modes
which arise only for the non-Abelian string (at $m_1=m_2$).
They are superpartners of the bosonic orientational moduli $n^a$;
therefore,  we will refer to these modes as superorientational.

To obtain these modes in the simplest and fastest way
we designed a special strategy which is outlined below.
We start from  the supersymmetry transformations for
the fermion fields in  the microscopic theory,
\begin{eqnarray}
\delta\lambda^{f\alpha}
&=&
\frac12(\sigma_\mu\bar{\sigma}_\nu\epsilon^f)^\alpha
F_{\mu\nu}+\epsilon^{\alpha p}D^m(\tau^m)^f_p\ +\dots,
\nonumber\\[3mm]
\delta\lambda^{af\alpha}
&=&
\frac12(\sigma_\mu\bar{\sigma}_\nu\epsilon^f)^\alpha
F^a_{\mu\nu}+\epsilon^{\alpha p}D^{am}(\tau^m)^f_p\ +\dots,
\nonumber\\[3mm]
\delta\bar{\tilde\psi}_{\dot{\alpha}}^{kA}
&=&
i\sqrt2\
\bar\nabla\hspace{-0.65em}/_{\dot{\alpha}\alpha}q_f^{kA}\epsilon^{\alpha
f}\ +\cdots,
\nonumber\\[4mm]
\delta\bar\psi_{\dot{\alpha}Ak}
&=&
 i\sqrt2\
\bar\nabla\hspace{-0.65em}/_{\dot{\alpha}\alpha}\bar
q_{fAk}\epsilon^{\alpha f}\ +\cdots.
\label{transf}
\end{eqnarray}
Here $\lambda^{f\alpha}$  and $\lambda^{af\alpha}$
are the fermions from the ${\cal N}=2$ vector supermultiplets of
the U(1) and SU(2) factors,  respectively,
 while $\psi^{kA}$ and $\tilde\psi_{Ak}$ are the fermion
partners of the squark fields $q^{kA}$ and $\tilde q_{Ak}$ in the
quark hypermultiplets. Moreover,
$f=1,2$ is the SU(2)$_R$ index,  $\alpha=1,2$ is the spinor
index,  $k=1,2$ is the color index, and $A=1,2$ is the flavor index.
The parameters of SUSY transformations
in the microscopic theory are denoted as  $\epsilon^{\alpha f}$.
Furthermore, the $D$ terms in Eq.~(\ref{transf}) are
\beq
D^1+ iD^2 =
i\, \frac{g_1^2}{2}\, \left({\rm Tr} \, |\Phi|^2-2\xi\right)\, ,\qquad D^3=0
\label{pdterm}
\eeq
for the U(1) field, and
\beq
D^{a1}+iD^{a2}=
i\, \frac{g_2^2}{2}\,{\rm Tr}\, \left(\Phi^{\dagger}\tau^a\Phi\right)\,,
\qquad
D^{a3}=0
\label{vdterm}
\eeq
for the SU(2) field. The dots in (\ref{transf})  stand for terms involving
the adjoint scalar fields which vanish on the string solution
(at $m_1=m_2$) because the
adjoint fields are given by their vacuum expectation values
(\ref{avevsu2}).

In Ref.~\cite{VY} it was shown that the four supercharges selected
by the conditions
\beq
\epsilon^{12}=-\epsilon^{11}\,, \qquad
\epsilon^{21}=\epsilon^{22}
\label{trivQ}
\eeq
act trivially on the BPS string. Now, to generate the superorientational
fermion zero modes we use  the following method. We assume that the
orientational moduli  $n^a$  in the string solution
(\ref{sna}) have a slow dependence on the
world-sheet coordinates $x_0$ and $x_3$ (or $t$ and $z$), as
in Sect.~\ref{kineticterm}. Then the four supercharges selected by
the conditions (\ref{trivQ}) no longer  act trivially. Instead,
their action now gives fermion fields proportional to the
$x_0$ and $x_3$ derivatives of $n^a$.
This is exactly what one expects from the residual \ntwo supersymmetry
in the world-sheet  theory.
The above four supercharges
generate the world-sheet supersymmetry in the \ntwo
two-dimensional $CP^1$ model. We use this world-sheet supersymmetry to
re-express the fermion fields obtained upon the action of these four
supercharges
in terms of the (1+1)-dimensional fermions. This will give us the
superorientational
fermion zero modes.

After this brief outline we can proceed to the implementation of
the  procedure. We substitute the
string solution (\ref{sna}) in (\ref{transf}) assuming that
$\epsilon^{\alpha f}$ are subject to the constraints (\ref{trivQ})
and the moduli $n^a$ have a slow dependence on the world-sheet coordinates.
Then we get
\beqn
\bar{\psi}_{Ak\dot{2}}
& =&
i\left(\tau^a\right)_{Ak}\left\{
\left[ (\pt_0-i\pt_3)n^a\right] \, (\phi_1-\phi_2)\,
\left( 1-\frac{\rho}{2}\right)
\right.
\nonumber\\[3mm]
&-& \left.
i\varepsilon^{abc}\, n^b\left[ (\pt_0-i\pt_3)\, n^c\right] \,
 \frac{\rho}{2}\,  (\phi_1+\phi_2)
\right\}\epsilon^{22}\,,
\nonumber\\[3mm]
\bar{\tilde{\psi}}^{kA}_{\dot{1}}
& =&
i\left(\tau^a\right)^{kA} \left\{
\left[ (\pt_0+i\pt_3)\, n^a\right] (\phi_1-\phi_2) \left(1-
\frac{\rho}{2}\right)\right.
\nonumber\\[3mm]
&+& \left.
i\varepsilon^{abc}\, n^b\, \left[(\pt_0+i\pt_3)\, n^c\right] \,
\frac{\rho}{2}\,
 \left(\phi_1+\phi_2\right)
\right\} \epsilon^{11}\, ,
\nonumber\\[3mm]
\bar{\psi}_{Ak\dot{1}}& =& 0\, ,\qquad
\bar{\tilde{\psi}}^{kA}_{\dot{2}}=0\, ,
\nonumber\\[4mm]
\lambda^{a22}
& =&
-\frac{x_1+ix_2}{r^2}
\left\{
\left[(\pt_0+i\pt_3)n^a\right] f_3\, (1-\rho)
+ i\varepsilon^{abc}\, n^b \left[(\pt_0+i\pt_3)n^c\right] r\frac{d}{dr}\rho
 \right\} \epsilon^{11}\, ,
\nonumber\\[3mm]
\lambda^{a11}
& =&
-\frac{x_1-ix_2}{r^2}
\left\{
\left[(\pt_0-i\pt_3)n^a\right] f_3\, (1-\rho)
-i\varepsilon^{abc}\, n^b \left[ (\pt_0-i\pt_3)n^c\right]
r\frac{d}{dr}\rho
 \right\}\epsilon^{22},
\nonumber\\[4mm]
\lambda^{12}
& =&
 \lambda^{11} \, ,\qquad  \lambda^{21}= -\lambda^{22}\,.
\label{zmodesdn}
\eeqn
Let us compare these transformations with the supersymmetry
transformations  on the string world sheet. For the model (\ref{o3n2}) they
are
\beq
\delta \chi^a=i\, \sqrt{2}\left[
 \left(\gamma_k \pt_k\ n^a\right)  \varepsilon  +
\varepsilon^{abc}\, n^b \left(\gamma_k \pt_k\ n^c\right)\eta
 \right] \,,
\label{susy2}
\eeq
where $\varepsilon^{\alpha}$ is a real two-component
parameter of the   two-dimensional supersymmetry, while  $\eta^{\alpha}$
is another such parameter ($\alpha=1,2$), so that assembled
together they form
a full set of  \ntwo transformations.
If we rewrite Eq.~(\ref{susy2}) in components,
\beqn
\delta \chi^a_1
&=&
i\sqrt{2} \left[\left( \pt_0 +i\pt_3\right) n^a \, \varepsilon_2
+\varepsilon^{abc}n^b\left( \pt_0 +i\pt_3\right) n^c\,\eta_2\right]\,,
\nonumber\\[3mm]
\delta \chi^a_2
&=&
i\sqrt{2} \left[\left( \pt_0 -i\pt_3\right)  n^a \, \varepsilon_1
+\varepsilon^{abc}n^b\left( \pt_0 -i\pt_3\right) n^c\,\eta_1\right]\, ,
\label{susy2c}
\eeqn
and identify properly normalized  parameters of the four-dimensional SUSY
transformations (with the  constraint (\ref{trivQ})) in
terms of $\varepsilon_{1,2}$, $\eta_{1,2}$, namely,
\beqn
\varepsilon_2 +i\eta_2
&=&
\frac1{\sqrt{2}} \left( \epsilon^{11}-\epsilon^{12}\right) =
\sqrt{2}\, \epsilon^{11}\, ,
\nonumber\\[3mm]
\varepsilon_1 -i\eta_1
&=&
\frac1{\sqrt{2}}\left( \epsilon^{22}+\epsilon^{21}\right) =
\sqrt{2}\, \epsilon^{22}\,,
\label{epsilon24}
\eeqn
we can express the derivatives of $n^a$
in Eq.~(\ref{zmodesdn})
in terms of $\chi_1^a$ and $\chi_2^a$, thus obtaining
the zero modes of the quark and gluino fields in terms of
the four superorientational moduli.  In this way we arrive at
\beqn
\bar{\psi}_{Ak\dot{2}}
& = &
\left(\frac{\tau^a}{2}\right)_{Ak}
\frac1{2\phi_2}(\phi_1^2-\phi_2^2)
\left[
 \chi_2^a
+i\varepsilon^{abc}\, n^b\, \chi^c_2\,
\right]\, ,
\nonumber\\[3mm]
\bar{\tilde{\psi}}^{kA}_{\dot{1}}
& = &
\left(\frac{\tau^a}{2}\right)^{kA}
\frac1{2\phi_2}(\phi_1^2-\phi_2^2)
\left[
 \chi_1^a
-i\varepsilon^{abc}\, n^b\, \chi^c_1\,
\right]\, ,
\nonumber\\[5mm]
\bar{\psi}_{Ak\dot{1}}
& = &
0\, , \qquad
\bar{\tilde{\psi}}^{kA}_{\dot{2}}= 0\, ,
\nonumber\\[4mm]
\lambda^{a22}
& = &
\frac{i}{2}\frac{x_1+ix_2}{r^2}
f_3\frac{\phi_1}{\phi_2}
\left[
 \chi^a_1
-i\varepsilon^{abc}\, n^b\, \chi^c_1
 \right]\, ,
\nonumber\\[4mm]
\lambda^{a11}
& = &
\frac{i}{2}\frac{x_1-ix_2}{r^2}
f_3\frac{\phi_1}{\phi_2}
\left[
\chi^a_2
+i\varepsilon^{abc}\, n^b\, \chi^c_2
 \right]\,,
\nonumber\\[4mm]
\lambda^{12}
& = &
 \lambda^{11} \, ,\qquad  \lambda^{21}= - \lambda^{22}\,,
\label{zmodes}
\eeqn
where we use the solution (\ref{rhosol}) for the function $\rho$
to simplify the expressions for the profile functions
of the fermion zero modes.
Equation (\ref{zmodes}) is our final result for the superorientational
fermion zero modes.  Here the dependence on $x_i$ is encoded in the
profile functions of the string, while $\chi^a_{\alpha}$ should be
considered as  {\em  constant Grassmann collective coordinates}.

To conclude this section let us check\,\footnote{For simplicity,
we restrict ourselves
to terms in the action quadratic in the fermion fields.} that
the zero modes above
do produce
the fermion part of the   \ntwo two-dimensional $CP^1$ model (\ref{o3n2}).
To this end we return to the usual assumption that
the fermion collective coordinates  $\chi^a_{\alpha}$ in
Eq.~(\ref{zmodes}) have an adiabatic  dependence on the world sheet
coordinates $x_k$ ($k=0,3$). This is quite similar to the procedure of
Sect.~\ref{kineticterm}.
Substituting Eq.~(\ref{zmodes}) in  the kinetic terms of fermions
in the microscopic theory,
\beq
\int d^4 x\left\{\bar{\psi}_A i\bar\nabla\hspace{-0.65em}/ \psi^A
+ \bar{\tilde{\psi}}^A i\bar\nabla\hspace{-0.65em}/ \tilde{\psi}_A
+\frac{i}{g_2^2}\bar{\lambda}_f^a \bar{D}\hspace{-0.65em}/\lambda^{af}
\right\}\,,
\label{fkin4}
\eeq
and taking into account the derivatives of  $\chi^a_{\alpha}$ with
respect to the world-sheet coordinates, after some algebra  we arrive at
\beq
\beta \int d^2 x \left\{\frac12 \, \bar{\chi}^a i\, \gamma_n
\pt_n\, \chi^a\right\},
\label{fkin}
\eeq
where $\beta$ is given by the same integral (\ref{beta}) as for
the bosonic kinetic term, see Eq.~(\ref{o3}). We see that (\ref{fkin})
exactly reproduces the kinetic term of the  (1+1)-dimensional
fermions in the $CP^1$ model (\ref{o3n2}).

\section{Sigma-model kinks: monopoles of the microscopic theory}
\label{sigmamodelkinks}
\setcounter{equation}{0}

Thus, we concluded our consideration of non-Abelian strings
in the microscopic theory and derivation of the macroscopic
world-sheet theory they induce. Our task in this section is
the study and analysis of the BPS string junctions --- the confined
monopoles --- both in the microscopic and macroscopic theories

A general picture is best inferred from consideration in the quasiclassical
regime, namely, we  start from the case $\Delta m \neq 0 $ but
subject to the constraint (\ref{mxi}).  This limit was studied in
Ref.~\cite{Tong:2003pz}.  In the   $CP^1$ model with the twisted mass
there is a kink (a.k.a domain wall) interpolating between two vacua at
the north and south poles, see Fig.~\ref{f13one}.
Two distinct strings of the microscopic
theory are the two vacua of the macroscopic model, while
the confined monopole is the $CP^1$ kink
interpolating between them. We intend to back
up this qualitative statement by quantitative data.

The $CP^1$ kink solution is
easy to find in the explicit form. This is discussed in
Sect.~\ref{kinksinthequasi}, as well as the occurrence of an
``extra" collective
coordinate.  Kinks in the \ntwo two-dimensional sigma model
with the twisted mass are exhaustively studied
in the literature \cite{Dorey}. These kinks are
1/2 BPS saturated (i.e. preserve two
supercharges). The corresponding collective coordinates are
$z_0$ (a complexified position of the center) and two fermion moduli.
We will say more on why $z_0$ gets complexified.

Now, if we send $\Delta m \to 0$  we still have two vacua in the $CP^1$
model, as was explained in Sect.~\ref{unequalquark}, and do have kinks
interpolating between them. This kink is best seen in the mirror
description of the model. It interpolates between the two vacua of
sine-Gordon potential (\ref{sGpot}).
The kinks are counted in dimension-two supermultiplets.
The  Cecotti--Fendley--Intriligator--Vafa (CFIV)
index \cite{CFIV} tells us that, in fact, we have two distinct
kinks, albeit this number is not invariant
under the variation of the twisted mass, see Sect.~\ref{multiplic}.

The $\Delta m=0$ kink should   be interpreted as a ``highly bound"
monopole (it has multiplicity two as well)
which realizes a junction of two ``elementary" strings.
However, the quantum numbers of this monopole are no longer $(1,-1)$.
Similar to strings, it does not have definite Abelian charges in the
limit $\Delta m \to 0$.  It becomes a {\em bona fide}
non-Abelian monopole of SU(2) representing the junction
between two ``elementary"   non-Abelian strings
associated with two quantum vacua of $CP^1$ model (Fig.~\ref{twoabcd},
the right lower corner).

There are two features of the $\Delta m=0$ kinks that we can established
on general grounds. First, their BPS saturated nature and
Eqs.~(\ref{13twelve}),
(\ref{13thirteen}) tell us that the kink mass is equal to \cite{Losev}
\beq
M_{k}= \frac{1}{\pi} \, |\Delta \langle \zeta^{-2}
\bar\Psi_R  \Psi_L \rangle | = \frac{2}{\pi} \, \Lambda_{\,
CP(1)}\,.
\label{exhaust}
\eeq
Second, the size of the kink (in the $z$ direction) is of the order
$\sim \Lambda_{\, CP(1)}^{-1}$.

This means that the SU(2) monopole, although classically  massless
and infinitely spread in the limit $\Delta m\to 0$,
in fact acquires a small but finite mass and finite
size due to non-perturbative effects on the string world sheet.
In  this section we give a detailed quantitative evidence
in favor of our identification of the SU(2) monopole as the
$CP^1$ model kink. We start with the quasiclassical limit of
non-zero $\Delta m$ considered  by Tong \cite{Tong:2003pz}
and then eventually arrive at
the quantum limit $\Delta m \to 0$.

\subsection{Microscopic theory: first-order master equations}
\label{firstorder}

In this section we derive the first-order equations for the 1/4-BPS
junction of the (1,0) and (0,1) strings in the quasiclassical limit
\beq
\Lambda_{CP(1)}\ll \mu  \ll g_2\sqrt{\xi}\, ,
\qquad \mu = \Delta m \, .
\label{lmxi}
\eeq
In this limit $\mu$  is small enough so we can use our effective
low-energy description in terms of the $CP^1$ model (with the twisted
mass). On the other hand, $\mu$  is
much larger then the scale of $CP^1$ model, so the latter is in the weak
coupling regime which  allows one
to apply the quasiclassical treatment.

The geometry of our junction is shown on Fig.~\ref{oneab}b. Both
strings are stretched along the $z$ axis. We assume that the monopole sits
near the origin, the (0,1)-string is at negative $z$, while the
(1,0)-string is at positive $z$. The perpendicular plane is parametrized by
$x_1$ and   $x_2$. What is sought for is  a static solution
of the BPS equations, with all relevant fields  depending  only on $x_1$,
$x_2$ and $z$.

Ignoring the time variable
we can represent the energy functional of our theory (\ref{redqed})
as follows (Bogomolny representation \cite{B}):
\beqn
E
&=&
\int{d}^3 x   \left\{
\left[\frac1{\sqrt{2}g_2}F^{*a}_{3} +
\frac{g_2}{2\sqrt{2}}
\left(\bar{\vp}_A\tau^a \vp^A\right)
+ \frac1{g_2}D_3 a^a\right]^2
\right.
\nonumber\\[3mm]
&+&
\left[\frac1{\sqrt{2}g_1}F^{*}_{3} +
\frac{g_1 }{2\sqrt{2}}
\left(|\vp^A|^2-2\xi \right)
+ \frac1{g_1}\pt_3 a\right]^2
\nonumber\\[4mm]
&+&
\frac1{g_2^2}\left|\frac1{\sqrt{2}}(F_1^{*a}+iF_2^{*a})+(D_1+iD_2)a^a
\right|^2
\nonumber\\[4mm]
&+&
\frac1{g_1^2}\left|\frac1{\sqrt{2}}(F_1^{*}+iF_2^{*})+(\pt_1+i\pt_2)a
\right|^2
\nonumber\\[5mm]
&+&
 \left|\nabla_1 \,\vp^A +
i\nabla_2\, \vp^A\right|^2
\nonumber\\[4mm]
&+&
\left. \left|\nabla_3 \vp^{A}+\frac1{\sqrt{2}}\left(a^a\tau^a +a
 +\sqrt{2}m_{A}
\right)\vp^A\right|^2\right\}
\label{bogj}
\eeqn
plus surface terms. Following our conventions we assume
the  quark masses to be real implying that  the vacuum expectation values
of  the adjoint scalar fields are   real too.
The surface terms mentioned above are
\beq
\left.
E_{\rm surface}=
\xi  \int d^3 x F^{*}_3
+\sqrt{2}\, \xi \int d^2 x \, \langle a\rangle\right|^{z=\infty}_{z=-\infty}
-\sqrt{2}\, \frac{\langle a^3\rangle}{g_2^2}\int d S_n\,  F^{*3}_{n},
\label{surface}
\eeq
where the integral in the last term runs over a large two-dimensional
sphere
at $\vec x^{\, 2} \to \infty$. The first term on the right-hand side
is related to strings, the second to domain walls, while the third to
monopoles (string junctions).

The Bogomolny representation (\ref{bogj})
leads us to the following first-order equations:
\beqn
&& F^{*}_1+iF^{*}_2 + \sqrt{2}(\pt_1+i\pt_2)a=0\, ,
\nonumber\\[3mm]
&& F^{*a}_1+iF^{*a}_2 + \sqrt{2}(D_1+iD_2)a^a=0\, ,
\nonumber\\[3mm]
&& F^{*}_{3}+\frac{g_1^2}{2} \left(\left|
\varphi^{A}\right|^2-2\xi\right) +\sqrt{2}\, \pt_3 a =0\, ,
\nonumber\\[3mm]
&& F^{*a}_{3}+\frac{g_2^2}{2} \left(\bar{\vp}_{A}\tau^a \varphi^{A}\right)
+\sqrt{2}\, D_3 a^a =0\, ,
\nonumber\\[3mm]
&& \nabla_3 \vp^A =-\frac1{\sqrt{2}}\left(a^a\tau^a+a+\sqrt{2}m_A\right)
\vp^A\, ,
\nonumber\\[4mm]
&& (\nabla_1+i\nabla_2)\varphi^A=0\, .
\label{foej}
\eeqn
These are our {\it master equations}.
Once these equations are satisfied the energy of the BPS object is
given by Eq.~(\ref{surface}).

Let us discuss the central charges (the surface terms)
of the string, domain wall and
monopole in more detail.  Say, in the
string case, the three-dimensional integral in the first term in
Eq.~(\ref{surface}) gives the length of the string times its flux. In  the
wall case, the two-dimensional integral in the second term in (\ref{surface})
gives the area of the wall times its tension. Finally, in the
monopole case the integral in the last term
in Eq.~(\ref{surface}) gives the magnetic-field flux. This means that
the first-order master equations (\ref{foej}) can be used to study
{\em strings, domain walls,  monopoles and all their possible junctions}.

It is instructive to  check that  the wall, the string and the monopole
solutions, separately, satisfy these equations. For the domain wall  this
check was done in \cite{SYsu3wall} where we used these equations to
study the string-wall junctions. Let us consider the string solution. Then
the scalar fields $a$ and $a^a$ are  given by their VEV's.
The gauge flux is directed along the $z$ axis, so
that $ F^{*}_1=F^{*}_2= F^{*a}_1=F^{*a}_2=0$.
All fields depend only on the perpendicular
coordinates $x_1$ and $x_2$. As a result,  the first
two equations and the fifth  one  in (\ref{foej}) are trivially
satisfied. The third and the fourth  equations
reduce to the first two equations in Eq.~(\ref{F38}). The last equation
in (\ref{foej}) reduces to the last equation in (\ref{F38}).

Now, turn to the monopole solution. The 't Hooft-Polyakov monopole
equations \cite{thopo}  arise from those in Eq.~(\ref{foej}) in the limit
$\xi=0$. Then all quark fields vanish,  and Eq.~(\ref{foej}) reduces to
the standard first-order equations for the BPS 't~Hooft-Polyakov monopole,
\beq
F^{*a}_k + \sqrt{2}\, D_k\,  a^a=0 \, .
\label{monopole}
\eeq
The U(1) scalar field $a$ is given by its VEV while the U(1) gauge field
vanishes.

Now, Eq.~ (\ref{surface}) shows that the central charge
of the SU(2) monopole is determined by $\langle a^3\rangle$ which is
proportional to the quark mass difference. Thus, for the monopole
on the Coulomb branch (i.e. with $\xi $ vanishing) Eq.~ (\ref{surface})
yields
\beq
M_m=\frac{4\pi\mu }{g_2^2}\, .
\label{mmCb}
\eeq
This coincides, of course, with the  Seiberg-Witten result
\cite{SW1} in the weak coupling limit. As we will see shortly,
the same expression continues to hold even if $\mu\ll \sqrt\xi$
(provided that $\mu$ is still much larger than $\Lambda$).
An explanation will be given in Sect.~\ref{dopo6}.

We pause here to make a remark on the literature.
The Abelian version of the first-order equations (\ref{foej})
were   derived in Ref. \cite{Shifman:2002jm} where they were
used to find the 1/4 BPS-saturated solution for the wall-string  junction.
A subset of   non-Abelian equations  (\ref{foej})
in the  SU(2)$\times$ U(1) theory was derived in \cite{Tong:2003pz}
with the purpose of  studying
the junction of two elementary  strings (``a confined monopole")
at $\Delta m \neq 0$. We extensively exploit this construction too,
as a reference point, while  our  main  interest is the limit $\Delta m=0$.
Non-Abelian equations (\ref{foej}) were derived and extensively
used   in the recent analysis \cite{SYsu3wall} of the wall-string
junctions for non-Abelian strings  ending on a stack of domain walls.

\subsection{The string junction solution for $\sqrt\xi\gg \mu\gg \Lambda$}
\label{stringjunctionsolution}

Now we apply our master equations in order
to find  the junction of the (0,1) and (1,0) strings via the $(1,-1)$
monopole
(see Fig.~\ref{twoabcd}, the left lower corner)
in the quasiclassical limit. We will  show that the solution of
the BPS equations (\ref{foej}) of the four-dimensional  microscopic
theory is determined by the kink
solution in the two-dimensional  sigma model (\ref{o3mass}).

To this end   we will look for the solution of equations (\ref{foej}) in the
following {\it ansatz}. Assume that the solution for the string
junction is given, to the leading order in $\mu/\sqrt{\xi}$,
by the same string configuration (\ref{sna}), and (\ref{aa}) which we
dealt with previously,  in Sect.~\ref{unequalquark}
(in the case   $\mu\neq 0$),
\beqn
\Phi
&=&
U\left(
\begin{array}{cc}
\phi_1(r) & 0  \\[2mm]
0 &  \phi_2(r)
\end{array}\right)U^{-1} =\frac12
(\phi_1+\phi_2)+n^a\,\frac{\tau^a}{2} \left( \phi_1-\phi_2\right) ,
\nonumber\\[4mm]
A^a_{i}(x)
&=&
n^a \,\varepsilon_{ij}\,\frac{x_j}{r^2}\, f_3(r),
\nonumber\\[4mm]
A_{i}(x)
&=&
\varepsilon_{ij}\,\frac{x_j}{r^2}\, f(r)\,,
\nonumber\\[5mm]
A^a_3
&=&
-\,\varepsilon^{abc}\, n^b  \left(\pt_3 \, n^c\right) \rho(r)\, ,
\qquad A^a_0=0\,,
\nonumber\\[4mm]
a^a
&=&
-\frac{\mu}{\sqrt{2}}\left[\delta^{a3}(1-\rho) +
n^a n^3\rho\, \right],\qquad    a=-\sqrt{2}\, m\, ,
\label{monsol}
\eeqn
with $n^a$    slowly-varying
functions of $z$, to be determined below,  replacing the  constant
moduli vector $\vec n$.
The {\em ansatz}  for the
gauge potentials $A_3^a$ and $A_0^a$
follows from Eqs.~(\ref{An}) and (\ref{endan}).
As   we have the
(0,1)-string at $z\to-\infty$, the function $n^a(z)$ satisfy the
boundary condition
\beq
n^a(-\infty) = (0,0,-1)\,,
\label{-infty}
\eeq
while
\beq
n^a(\infty) = (0,0,1)\,.
\label{infty}
\eeq
The latter condition ensures that we have the (1,0)-string at $z\to\infty$.
The {\it ansatz} (\ref{monsol}) corresponds to the non-Abelian string
in which the vector $n^a$ slowly rotates from (\ref{-infty}) at
$z\to-\infty$  to  (\ref{infty}) at  $z\to\infty$. Now we will show that
the representation (\ref{monsol})  solves the master
equations (\ref{foej}) provided the functions $n^a (z)$ are chosen
in a special way.

Note that the first equation in (\ref{foej}) is trivially
satisfied because the field $a$ is  constant and $F^{*}_1=F^{*}_2=0$.
The last equation reduces to the first two  equations in (\ref{foest})
because it does not contain derivatives with respect to $z$
and, therefore, is satisfied for arbitrary functions $n^a(z)$.
The same remark applies also to the third equation in Eq.~(\ref{foej}),
which reduces to  the third equation in  (\ref{foest}).

Now let us consider the fifth equation in  Eq.~(\ref{foej}).
Substituting (\ref{monsol}) in this equation and using
expression  (\ref{rhosol}) for $\rho$  we find that this equation
is satisfied provided $n^a (z) $ are chosen to be the  solutions of the
equation
\beq
\pt_3 n^a=\mu \left(\delta^{a3}-n^a n^3\right)\, .
\label{kinkeq}
\eeq
This equation, written in the holomorphic representation, is discussed
in Sect.~\ref{kinksinthequasi}.

By the same token, we can consider the second equation in  (\ref{foej}).
Upon substituting there the {\em ansatz}  (\ref{monsol}),  it reduces to
Eq.~(\ref{kinkeq}) too.
Finally, consider the fourth equation in  (\ref{foej}). One can see that
in fact
it contains an
expansion in the parameter $\mu^2/\xi$. This means  that the solution
we have just built  is not exact; it has corrections
of the order of $O(\mu^2/\xi)$. To the leading order in this parameter
the fourth equation in (\ref{foej}) reduces to the last equation
in (\ref{foest}). In principle, one could go beyond
the leading order.
Solving the fourth equation in (\ref{foej}) in the next-to-leading
order would allows one  to determine $O(\mu^2/\xi)$
corrections to our solution (\ref{foest}). This goes beyond
the scope of our current investigation.

Let us dwell on the meaning of Eq. ~(\ref{kinkeq}). This equation
is nothing but
the equation for the kink in the $CP^1$ model (\ref{o3mass}).
A thorough analysis of the  $CP^1$-model kinks will be carried out in
Sect.~\ref{kinksinthequasi}. Here we will limit ourselves to
the Bogomolny completion of this model. The energy functional can be
rewritten as
\beq
E= \frac{\beta}{2}\int d z \left\{\left| \pt_z n^a-
\mu\left (\delta^{a3}-n^a n^3 \right)\right|^2
+2\mu\,  \pt_z n^3\right\}\,.
\label{bogcp1}
\eeq
The above  representation implies the first-order equation (\ref{kinkeq})
for the BPS saturated kink. It also yields $2\beta\mu$ for the kink mass.

Thus, we have demonstrated  that the junction solution
for the (0,1) and (1,0) strings is given by the non-Abelian string
with a slowly varying orientation vector $n^a$. The  variation of $n^a$
is described in terms of the kink solution of the (1+1)-dimensional
$CP^1$ model with the twisted mass. This  was
expected.

In conclusion, we would like to match the masses of the
four-dimensional mono\-pole and two-dimensional
kink. The string mass and that of the string junction is
given by first and the last
terms in the surface energy (\ref{surface}) (the second term
vanishes). The first term obviously reduces to
\beq M_{\rm string}=2\, \pi\, \xi\,\, L,
\label{stringm}
\eeq
i.e.  proportional to the total string length $L$. Note that
both the (0,1) and (1,0) strings have the same tension (\ref{pmst}).
The third term should give the mass of the (1,-1) monopole.
The surface integral in this term reduces
to the flux of the (1,0)-string  at $z\to\infty$ minus  the flux
of the (0,1)-string  at $z\to -\infty$. The $F^{*3} $ flux of the
(1,0)-string is $2\pi$ while the $F^{*3} $ flux of the (0,1)-string is
$-2\pi$. Thus, taking into account Eq.~(\ref{avev}), we get
\beq
M_m=\frac{4\pi}{g_2^2}\, \mu\, .
\label{mm}
\eeq
Note, that although  we discuss the monopole in the confinement
phase  at $\left| \Delta m\right| \ll \sqrt{\xi}$ (which is
a junction of two strings  in this phase), nevertheless
in terms of the $\mu$ and $g^2_2$
dependence  its mass coincides with the result (\ref{mmCb})
for the unconfined monopole on the Coulomb branch (i.e. at $\xi=0$).
There is no change in the monopole mass formula
in the first three cases in  Fig.~\ref{twoabcd}.
This is no accident --- there is a deep theoretical reason explaining the
validity of the unified formula. A change occurs only in passing to the
highly quantum regime depicted in the right
lower corner of Fig.~\ref{twoabcd}.  We will discuss this regime shortly,
see Eq.~(\ref{exhaus}), while more details will be given  in
Sect.~\ref{dopo6}.

Now let us compare (\ref{mm}) with the kink mass in the effective
$CP^1$ model on the string world sheet. As was mentioned,
the surface term in Eq.~(\ref{bogcp1}) gives
\beq
M_{\rm kink}=2\,\beta\, \mu = \frac{\mu}{\pi}\, \ln
\frac{\mu}{\Lambda_{2D}}\, ,
\label{exhau}
\eeq where in the second
equality we used Eq.~(\ref{sigmacoup}) to express $\beta$ in terms of
the dynamical scale parameter. Two remarks are in order here.  First,
expressing the two-dimensional coupling constant $\beta$ in terms of
coupling constant of the microscopic theory, see Eq.~(\ref{betagamma}),
we  obtain
\beq
M_{\rm kink}=\frac{4\pi}{g^2_2}\mu\, ,
\label{mk}
\eeq
thus verifying that the four-dimensional
calculation of $M_m$ and the two--dimensional
calculation of $M_{\rm kink}$ yield the same,
\beq
M_m=M_{\rm kink}\, .
\label{monkink}
\eeq
Needless to say, this is in full accordance with the physical picture
that emerged from our analysis, that the two-dimensional $CP^1$
model is nothing but the macroscopic description
of the confined monopoles occurring in the four-dimensional
microscopic Yang-Mills theory.

Technically the coincidence of the monopole and kink masses
is based on the fact that the integral in the
definition (\ref{beta}) of the sigma-model coupling $\beta$ is unity.

The second remark concerns the second equality in Eq.~(\ref{exhau}).
In fact, using this form, one can get \cite{Dorey} a unified formula
for $M_{\rm kink}$ (and, hence, for $M_m$) describing the  last two
regimes in Fig.~\ref{twoabcd}. To this end one replaces the logarithm
\beq
\ln \frac{\mu}{\Lambda_{2D}} \to \left| \frac{1}{2}\,
\ln\frac{\sqrt{\mu^2+4\Lambda_{\,CP(1)}^2}+\mu }{
 \sqrt{\mu^2+4\Lambda_{\,CP(1)}^2}-\mu} -
\sqrt{1+\frac{4\Lambda_{\,CP(1)}^2}{\mu^2}}\,
\right|\,.
\label{exhaus}
\eeq
In the quasiclassical regime $\mu /\Lambda \gg 1$
the right-hand side and left-hand
side coincide provided $\Lambda_{2D} = e\Lambda_{\,CP(1)}$.
If, on the other hand, $\mu\to 0$, combining Eqs.~(\ref{exhau})
and  (\ref{exhaus}) we arrive at Eq.~(\ref{exhaust}) for the kink
mass. Using our identification of the four-dimensional monopole
as a two-dimensional kink we then get the confined monopole mass,
\beq
M_m=\frac2{\pi}\Lambda_{CP(1)}\,,
\label{mmmu0}
\eeq
in the limit $\mu\to 0$.

\subsection{More on kinks in the $CP^1$ model with
the twisted mass}
\label{kinksinthequasi}

Identification of the confined monopoles in four dimensions
with the two-dimensional kinks yields an immediate bonus:
all we know of the $CP^1$-model kinks can be rephrased in
terms of the confined monopoles.
The goal of this section is
to briefly review $CP^1$-model kinks, with an eye on the
parallel with the confined monopoles. Introductory data on the
$CP^1$ model can be found in Sects.~\ref{11dimensional}
and \ref{unequalquark}. We will heavily rely on the results of
Refs.~\cite{Dorey,Losev}.
The reader familiar with these results can proceed directly to
Sect.~\ref{implic}.
Among other topics, we will dwell
on the role of the $\theta$ term and on
``dyonic" kinks which are counterparts of ``dyonic"
confined monopoles of the microscopic theory.

\subsubsection{The $\theta$ term}
\label{thetaterm}

So far we chose $\mu$ to be real, and put $\theta =0$.
Both requirements can and must be relaxed.
Let us parametrize $\mu$ as
\beq
\mu = |\mu |e^{i \omega}\,.
\label{parmu}
\eeq
The chiral anomaly implies that neither $\omega$ not $\theta$
are separately  observable. The physically observable phase combination is
\beq
\theta_{\rm eff} = \theta +2\omega\,.
\label{thetaeff}
\eeq
Correspondingly, we can always eliminate
$\theta$ by including it in the definition of the phase of $\mu$.
Alternatively, we can always define $\mu$ to be real and positive,
at a price of shifting the original $\theta$
by an appropriate amount, according to Eq.~(\ref{thetaeff}). Note that
$\theta$ is defined mod $2\pi \, k$ where $k$ is an integer.

\subsubsection{Superalgebra at $\mu\neq 0$}
\label{supalmunz}

In the $CP^1$ model with the  twisted mass
\beqn
\{Q_L \bar  Q_R\} &=&-i \, \mu q_{\rm U(1)}-  \mu \, \int
dz\, \partial_z\, h\, + \frac{ 1}{ \pi}
\int dz\, \partial_z \left(\zeta^{-2}\,
\bar\Psi_R  \Psi_L \right) \,,
\nonumber \\[2mm]
\{Q_R \bar
Q_L\}&=&i \,  \bar \mu q_{\rm U(1)} - \bar \mu \, \int dz\, \partial_z\, h
\, + \frac{ 1}{ \pi}  \int dz\, \partial_z \left(\zeta^{-2}\,
\bar\Psi_L  \Psi_R \right) \,,
\label{13twenty}
\eeqn
where $q_{\rm U(1)}$ is the conserved U(1) charge,
\beqn
q_{\rm U(1)}&\equiv& \int dz \, {\cal J}^0_{\rm U(1)}\,,
\nonumber\\[4mm]
{\cal J}^\mu_{\rm U(1)} &=& G \left( \bar w \,i\,
\stackrel{\leftrightarrow}{\partial}^{\, \mu}\! w  +
\bar\Psi \gamma^\mu\Psi -2\,\frac{w \bar w}{\zeta}\,
\bar\Psi \gamma^\mu\Psi
\right)\,,
\label{13twentyone}
\eeqn
and
\beq
h = -\frac{2}{g^2_0}\,\, \frac{1}{\zeta }\,,
\label{13twentytwo}
\eeq
where $g_0^2$ is the bare coupling constant of the $CP^1$ model
(i.e. normalized at an ultraviolet scale $M_{UV}$).
The supercharges are normalized in such a way
(see Eq.~(\ref{13thirteen})) that the kink mass
equals the absolute value of the expectation value of the right-hand
side. If $\mu$ is real,
\beq
q_{\rm U(1)} =\frac{\theta_{\rm eff}}{2\pi}\,,
\label{quuone}
\eeq
and, quasiclassically,
\beq
M_{\rm kink} = \left|\mu \left(
\frac{2}{g^2_0} -\frac{1}{2\pi}\ln\frac{M_{UV}^2}{\mu^2}
+i \frac{\theta_{\rm eff}}{2\pi}\right) \right|\,.
\label{kmtz}
\eeq
At $\theta_{\rm eff} =0$ Eqs.~(\ref{kmtz}) and (\ref{exhau}) are identical.

\subsubsection{Kinks in the quasiclassical limit}
\label{inthequasi}

In the quasiclassical limit (\ref{lmxi})
the physical meaning of the kink is absolutely
transparent. It is the tunneling trajectory from the north pole
(the (1,0)-string) to the south pole (the (0,1)-string).
The BPS  equation  in the
holomorphic representation reads
\beq
\partial_z\bar{w} = -\mu \bar{w}\quad{\rm or}\quad
\partial_z w =
-\mu  w\,,
\label{13twentysix}
\eeq
to be compared with Eq.~(\ref{kinkeq}). Here $\mu $
is assumed real and positive, see Sect.~\ref{thetaterm}.

\begin{figure}[h]
\epsfxsize=6cm
\centerline{\epsfbox{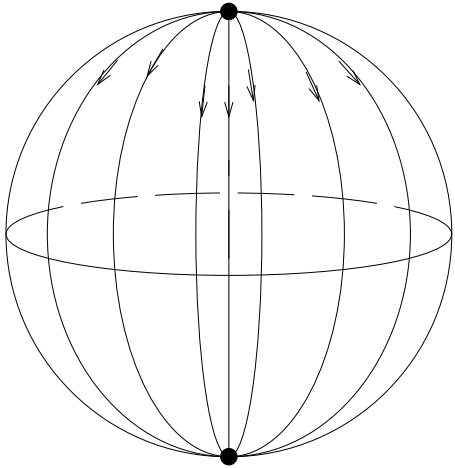}}
 \caption{The soliton solution family.
The collective coordinate $\alpha$ in Eq. (\ref{13twentyseven})
spans the interval $0\leq\alpha\leq 2\pi$. For given $\alpha$
the soliton trajectory on the target space sphere
follows a   meridian,
so that when $\alpha$ varies from 0 to $2\pi$
all meridians are covered.}
 \label{f13two}
\end{figure}

The BPS equation (\ref{13twentysix}) has a number peculiarities
the most important of which is its complexification,
i.e. the fact that Eq. (\ref{13twentysix}) is holomorphic in $w$.
The solution of this equation is, of course, trivial, and can be written
as
\beq
w (z) = e^{-\mu\, (z-z_0) -i\alpha}\,.
\label{13twentyseven}
\eeq
Here $z_0$ is the kink center while $\alpha$ is an arbitrary phase.
In fact, these two parameters enter only in the combination
$\mu\,  z_0 -i\alpha$. As was noted, the notion of the kink center
 gets complexified.
The physical meaning of the modulus $\alpha$ is obvious:
there is a continuous family of solitons
interpolating between the north and south poles of the target space sphere.
This is due to U(1) symmetry.
The soliton trajectory can follow any meridian (Fig.~\ref{f13two}).
As we will see shortly,
there are two fermion counterparts of $z_0$ and $\alpha$,
which will be referred to as $\eta$ and $\bar\eta$.

\subsubsection{Quantization of the bosonic moduli}
\label{quantibosmo}

To carry out  conventional
quasiclassical quantization we, as usual,
assume the moduli $z_0$ and $\alpha$ in Eq. (\ref{13twentyseven})
to be (weakly) time-dependent, substitute  (\ref{13twentyseven})
in the bosonic Lagrangian (\ref{o3massw}), integrate over $z$
and thus derive
a quantum-mechanical Lagrangian describing moduli dynamics.
In this way we obtain
\beq
{\cal L}_{\rm QM} = -M_{\rm kink} +\frac{M_{\rm kink}}{2} \, \dot z_0^2
+ \frac{\beta}{\mu }\,\,  \dot \alpha ^2  -\frac{\theta_{\rm eff}}{2\pi}\,
\dot \alpha\,.
\label{13twentynine}
\eeq
The variable $\alpha $ is compact. The canonic momentum $\pi_\alpha$
conjugate to
$\alpha$ must be defined as
\beq
\pi_\alpha = \frac{2\beta}{\mu}\, \dot \alpha-\frac{\theta_{\rm eff}}{2\pi}
\longrightarrow  -i \frac{d}{d\alpha }\,.
\label{canmca}
\eeq
With these definitions
\beq
q_{\rm U(1)} = \frac{\theta_{\rm eff}}{2\pi} +\pi_\alpha\,,
\label{qccha}
\eeq
while the Hamiltonian
\beq
H = M_{\rm kink} + \frac{\mu}{4\beta}
\left(\frac{\theta_{\rm eff}}{2\pi}\right)^2 +
O(\pi_\alpha,\,\,\pi^2_\alpha)\,.
\label{qmene}
\eeq
For the BPS state $\pi_\alpha =0$, and Eqs.~(\ref{qccha}) and (\ref{qmene})
are consistent with Eq.~(\ref{kmtz}). In fact, they describe a whole
tower of BPS kinks since $\theta_{\rm eff}$ is defined modulo
$2\pi \, k$ with integer $k$.
The states with $q_{\rm U(1)}\neq 0$ are ``dyonic" kinks.
They have a nonvanishing charge with respect to the global U(1)
symmetry present in the CP(1) model with the twisted mass.

\subsubsection{Switching on fermion moduli}

The equations for the fermion zero modes of the kink are
\beqn
&\partial_z\Psi_L&    - \frac{2}{\zeta}\left(\bar w\partial_z w
\right)\Psi_L  + \frac{1-\bar ww}{\zeta}\, \mu \, \Psi_R
=0\,,
\nonumber\\[3mm]
&\partial_z\Psi_R&   - \frac{2}{\zeta}\left(\bar w \partial_z w
\right)\Psi_R -  \frac{1-\bar ww}{\zeta}\, \mu \,\Psi_L=0\,,
\label{13fourtyfour}
\eeqn
plus   similar equations for $\bar\Psi$.
It is not difficult to find normalizable solutions to these
equations, either directly, or using supersymmetry,
\beq
\left(\begin{array}{c}
\Psi_R\\  \Psi_L
\end{array}
\right)=\eta
\left(\frac{\mu }{2\beta}\right)^{1/2}
\left(\begin{array}{c}
-i\\   1
\end{array}
\right)e^{- \mu (z-z_0)}
\label{13fourtyfive}
\eeq
 and
 \beq
\left(\begin{array}{c}
\bar \Psi_R\\  \bar \Psi_L
\end{array}
\right)=\bar\eta \left(\frac{\mu }{2\beta}\right)^{1/2}
\left(\begin{array}{c}
i\\   1
\end{array}
\right)e^{- \mu (z-z_0)}\,,
\label{13fourtysix}
\eeq
where the numerical prefactors are introduced to ensure proper
normalization of the quantum-mechanical Lagrangian.

 Now, to
 perform   quasiclassical quantization
the fermion moduli $\eta$, $\bar\eta$ are assumed to be
time-dependent, and we derive their
quantum mechanics  starting from the original Lagrangian
(\ref{13one}), (\ref{13sixteen}),
\beq
{\cal L}'_{\rm QM} = i\, \bar\eta\dot\eta\,,
\label{13fourtyseven}
\eeq
implying the anticommutation relations
\beq
\{\bar\eta\eta\} =1\,,\quad \{\bar\eta \bar\eta\} =0\,,
\quad \{\eta \eta\} =0\,,
\label{13fourtyeight}
\eeq
which tell us that the wave function is {\em two-component}
(i.e. the kink supermultiplet is two-dimensional). One can implement
Eq. (\ref{13fourtyeight}) by choosing e.g.
$\bar\eta=\sigma^+$, $ \eta=\sigma^-$. The eigenstates then will be of
the type $|\uparrow\rangle$ and $|\downarrow\rangle$.

Upon quantization of the fermion moduli
one finds that the U(1) charge
of the BPS kink states gets a fractional shift,
\beq
\frac{\theta_{\rm eff}}{2\pi} \to \frac{\theta_{\rm eff}}{2\pi}\,
\pm \,  \frac{1}{2}\,,
\label{fracshift}
\eeq
where the plus or minus signs correspond to
 $|\uparrow\rangle$ and $|\downarrow\rangle$, respectively.
This fractional shift is due to fact that
there are two fermion zero modes, and is conceptually similar
to the well-known charge fractionalization phenomenon \cite{jackiwr}.

\subsubsection{Exact solution}
\label{exactsoln}

The above features of the BPS kinks in the
$CP^1$ model are concisely summarized
by the exact expression for the corresponding central charge
(cf. Eq.~(\ref{13twenty}))
\beq
Z_{2D} = i\, \mu \, q + \mu_D\, T
\label{nickone}
\eeq
obtained in Ref.~\cite{Dorey} exploiting methods similar to those of
Seiberg and Witten \cite{SW1}. Here $T$ is the topological charge of the
kink under consideration. In this work
\beq
T=1\,,
\label{topoone}
\eeq
while the parameter $q$ in Eq.~(\ref{nickone})
\beq
q = 0,\,\,\pm 1,\,\, \pm 2\,, ...
\label{qez}
\eeq
The quantity
$\mu_D$ is introduced in analogy with $a_D$ of Ref.~\cite{SW1},
\beq
\mu_D =\frac{\mu}{\pi}\left[
\frac{1}{2}\,
\ln\frac{\mu  + \sqrt{\mu^2+4\Lambda_{\,CP(1)}^2e^{-i\theta}}}{\mu -
 \sqrt{\mu^2+4\Lambda_{\,CP(1)}^2e^{-i\theta}}} -
\sqrt{1+\frac{4\Lambda_{\,CP(1)}^2e^{-i\theta} }{\mu^2}}\,\, \right]\,,
\label{nicktwo}
\eeq
where $\mu$ is now assumed to be complex,
as in Eq.~(\ref{parmu}).
The two-dimensional central charge is normalized in such a way that
$M_{\rm kink} = |Z_{2D}|$.

Note that the integer parameter $q$ in Eq.~(\ref{nickone}) is {\em not}
the physical U(1) charge. The latter is related to $q$ is follows (at $T=1$):
\beq
q_{\rm U(1)} = q +{\rm Im}\, \frac{\mu_D}{\mu}\,.
\label{physuone}
\eeq

The limit $|\mu | /\Lambda_{\,CP(1)} \to \infty$
corresponds to the quasiclassical domain, while corrections
of the type $(\Lambda_{\,CP(1)}/\mu)^{2k}$ are induced by instantons.

\subsubsection{Multiplicity}
\label{multiplic}

The two-dimensional $CP^1$ model
has  ${\cal N} =2$ supersymmetry; correspondingly, each shortened
supermultiplet is two-dimensional. In fact, one can introduce a
fermion parity \cite{Losev},
and each shortened supermultiplet has one plus and one minus state
with respect to this parity. The question to be addressed  below is
whether one has an extra degeneracy, and -- if yes -- how many
degenerate two-dimensional supermultiplets one has for given values
of  parameters. We will limit ourselves to
$T=1$, but the value of $q_{\rm U(1)}$ can be arbitrary.

Let us start from the
quasiclassical limit $\mu \gg \Lambda_{CP(1)}$ assuming $\mu$ to be real.
Of course, at $\mu\neq 0$, the results are $\theta$-dependent.
Let us consider a general case,  $\theta\neq 0$. Then,
for each value of the soliton mass
determined from  Eqs.~(\ref{nickone}) and  (\ref{nicktwo})
we have one two-dimensional supermultiplet.
There is a whole tower of  $T=1$  solitons
corresponding to $q = k$.
In this tower a single two-dimensional supermultiplet is the lightest.

Now, if  $\theta = 0$ (or $\pm 2\pi$, $\pm 4\pi$, etc.), we get a  a
special case, because in this case
the states with distinct $k$ conspire. For each given value of the mass
from  Eqs.~(\ref{nickone}) and  (\ref{nicktwo}) we have two degenerate
two-dimensional supermultiplets.

One can readily  rephrase these statements allowing oneself
to travel in the complex $\mu$ plane.
The above  degeneracy will hold provided $\theta_{\rm eff} =0$.

What happens when one travels from the domain of large $|\mu |$ to that
of small $|\mu |$?
If $\mu=0$ we know e.g.  from the mirror representation (\ref{sG}),
that there are two degenerate two-dimensional supermultiplets,
corresponding to the CFIV index = 2. Of course, at $\mu =0$
there is no $\theta$ dependence, and two BPS supermultiplets --  those with $\left\{q,\,\, T\right\} $  charges (0, 1) and (1,1) -- are degenerate.
Away from the point $\mu = 0$ the masses of these states are no longer equal; there are two singular points with one of
the two states becoming massless
at each. The region containing the point $\mu = 0$
is separated from the quasiclassical region of large $\mu$
by an infinite family of curves of the marginal stability (CMS)
on which the infinite number of other BPS states, visible
quasiclassically, decay.
Thus, the infinite tower of the $\left\{q,\,\, T\right\} $
BPS states existing in the quasiclassical domain degenerates in
just two stable BPS states in the vicinity of $\mu = 0$.

The CFIV index for the BPS states in question is independent of  $D$
and $F$ terms
but does depend, generally speaking,  on twisted $F$ terms,  and the
twisted mass parameter $\mu$,  in particular.  The  CFIV index  can
change discontinuously as one crosses a CMS. For more details see e.g.
the last work in Ref.~\cite{CFIV}
(in particular, Eq. (2.9))  although this work does not allow for the
possibility of a global Noether charge such as  $q_{\rm U(1)}$.

In the four-dimensional ${\cal  N}=2$ SUSY theories
on the Coulomb branch it turned out possible \cite{BF}
to find all  CMS explicitly by a careful  study of the phase of
the exact central charges. Since the latter are
fully equivalent  \cite{Dorey} to a central charge in the two-dimensional
sigma model,
a similar analysis should go through in the $CP^1$ model as well, see also
Sect.~\ref{dopo6}.

\subsection{Implications for confined monopoles}
\label{implic}

Since, as we have proven,  the $CP^1$ model presents
the macroscopic description of solitons in our four-dimensional microscopic model, all results regarding the BPS kinks summarized in
Sect.~\ref{kinksinthequasi}
can be immediately translated in statements regarding the confined
monopoles -- 1/4 BPS states in ${\cal N}=2$ four-dimensional Yang-Mills theory.

If $|\Delta m |\gg \Lambda$, our microscopic model has an infinite
tower of  ``dyonic" confined monopoles. In addition to the topological charge $T$ these BPS states carry a  Noether U(1) charge. This U(1) charge has nothing to do with the electric charge of the Julia-Zee dyons \cite{JZ}. The latter is associated with the gauge U(1) symmetry which remains unbroken in the 't Hooft-Polyakov
theory.  In our theory the gauge symmetry is completely
broken; there are no long-range forces. It is a {\em global} U(1) symmetry
that survives. The ``dyonic" confined monopoles are charged
with respect to this global U(1). Their masses are given by Eqs.~(\ref{nickone}),  (\ref{nicktwo}), with $T=1$ and $q=k$,
see also (\ref{physuone}). Generally speaking, the ``dyonic"
confined monopoles carry irrational U(1) charges.
Thus, in the presence of the $\theta$ term, they
experience the same charge ``irrationalization"
\cite{wph} as the   't Hooft-Polyakov monopoles (the
Witten phenomenon).\footnote{ For semi-integer
$q_{\rm U(1)}$ there is an additional mass degeneracy.}

Needless to say, the dyonic confined monopoles do not exist
without the non-Abelian strings attached to them.
The latter carry a non-Abelian magnetic flux. The elementary
excitations of these strings carry integer $q_{\rm U(1)}\neq 0$ but $T=0$.
As $|\Delta m |$ decreases, the dyonic confined monopoles
become unstable, as one passes through a family of CMS.
Eventually, only two monopole supermultiplets survive
as stable 1/4 BPS saturated
states.  At $\Delta m =0$ they are degenerate,  which reflects
the global SU(2) symmetry of the microscopic model.

At $\Delta m \neq 0$
the nontopological (i.e. $T=0$ and $q_{\rm U(1)}=\pm 1$) excitations
of the string are BPS states with mass $\mu$ confined to the string.
They can be interpreted
as follows.  Inside the string the squark profiles
vanish, effectively bringing us towards the Coulomb branch ($\xi =0$)
where the $W$ bosons and 
quarks would become BPS saturated
states in the bulk. As a matter of fact,
on the Coulomb branch the $W$ boson  
and off-diagonal quark mass would just equal $\mu$.
Hence,   the  $T=0$ BPS excitation
of the string is a wave of such $W$ bosons/quarks propagating along the 
string.  One could   call it a ``confined $W$ boson/quark." It is 
localized in the perpendicular but not in the transverse direction.  
What is important, it has no connection with the bulk Higgs phase $W$ 
bosons which are not BPS and much heavier than $\mu$. Neither these 
nontopological excitations have connection with the bulk quarks in our 
microscopic model which are not BPS saturated too.

\section{Anomaly: matching the central charges}
\label{quantumlimit}
\setcounter{equation}{0}

If $\mu =0$, which will be assumed in this section,
the $CP^1$ model runs into strong coupling and its physics is
determined by quantum effects. In particular, at $\mu = 0$
the vacuum expectation value
 $\langle a^3 \rangle =0$; the  strings become genuinely
non-Abelian.

There are no massless states in the $CP^1$ model
at $\mu =0$.
In particular, the kink mass   is of  the order of $\Lambda_{CP(1)}$,
as it is clear e.g. from the mirror description of the model (\ref{sG}).
On the other hand, in this limit both the last term in (\ref{surface})
and the surface term in (\ref{bogcp1}) vanish for
the monopole and kink masses, respectively.
This puzzle is solved by the following observation:
anomalous terms  in the central charges of   both four-dimensional
and two-dimensional SUSY algebras emerge.
Below we  discuss   the relation between the central charge anomalies in the
microscopic and macroscopic theories.

In the microscopic
theory the central charge associated with the monopole
has the following general form:
\beq
\{Q_\alpha^f\,  Q_{\ \beta}^g\} =\varepsilon_{\alpha\beta}\,
\varepsilon^{fg}\,2\,  Z_{4D}\,,
\eeq
where $Z_{4D}$ is an SU(2)$_R$ singlet. It is most convenient to write
$Z_{4D}$ as a topological charge (i.e. the integral over a topological
density),
\beq
Z_{4D} = \int \, d^3 x \, \zeta^0 (x)\,.
\label{4dcc}
\eeq
In the model at hand
\beqn
\zeta^\mu
&=&
\frac12\varepsilon^{\mu\nu\rho\sigma}
\,\partial_\nu\left(
\frac{i}{g_2^2}\, a^a F^a_{\rho\sigma} + \frac{i}{g_1^2}\, a F_{\rho\sigma}
\right.
\nonumber\\[4mm]
&+&
\left.
\frac{c}{4\pi^2}\left[
\lambda^a_{f\alpha}(\sigma_{\rho})^{\alpha\dot\alpha}
(\bar{\sigma}_{\sigma})_{\dot\alpha\beta} \lambda^{af\beta}+
2g_2^2\tilde\psi_{A\alpha}(\sigma_{\rho})^{\alpha\dot\alpha}
(\bar{\sigma}_{\sigma})_{\dot\alpha\beta}\psi^{A\,\beta}
\right]
\right)\,,
\label{4anom}
\eeqn
where $c$ is a numerical coefficient
which can be obtained from a one-loop calculation in the
${\cal N}=2$ regularized SU(2)$\times$U(1) gauge theory
which is our microscopic model. The operator in the
square brackets represents the anomaly, since $c$ vanishes at the tree
level. Note that the  general structure of the operator in the
square brackets is unambiguously fixed by
dimensional arguments, the Lorentz symmetry and other symmetries
of the microscopic theory. It is only the coefficient $c$ which is unknown.
The anomalous term plays a crucial role in the Higgs phase.\footnote{
A similar (albeit distinct) effect exists  on the Coulomb branch.  
The relationship between the 't~Hooft-Polyakov monopole mass
and the ${\cal N}=2$ central charge is analyzed
in the recent publication \cite{Rebhan:2004vn}, which identifies
an anomaly in the central charge explaining  
a constant (i.e. non-logarithmic) term in the monopole mass
on the Coulomb branch. The result of Ref.~\cite{Rebhan:2004vn} is in
agreement with the  Seiberg-Witten formula for the monopole mass.}

It is quite difficult to calculate $c$ directly because to this end one
needs
an explicit ${\cal N}=2$ ultraviolet regularization of the
four-dimensional theory. Although this is doable, the direct
calculation has not been completed yet. However, one can find $c$
indirectly, by comparing the expressions
for the masses of the BPS saturated confined monopole on the one hand, and
the BPS saturated kink in the $CP(1)$ on the other hand.
More precisely, we will compare the corresponding central charges
in the microscopic and macroscopic theories.

The mass of the monopole in terms of the central charge is given by
\beq
M_m=\sqrt{2}\left |Z_{4D}\right |\,.
\label{mz}
\eeq
In the limit $\mu =  0$ the classical term in $\left |Z_{4D}\right |$
vanishes, and
the central charge is determined by the last anomalous term in
Eq.~(\ref{4anom}). Now our task is to project it onto the
macroscopic theory.
Substituting the superorientational fermion  zero modes
of the string
(\ref{zmodes}) in the square brackets in Eq.~(\ref{4anom}) we get
\beqn
Z_{4D}
&=&
-\frac{c}{2\pi}\int d z \pt_z\left(\chi_1^a\chi_2^a
-i\varepsilon^{abc}n^a\chi_1^b\chi_2^c\right)
\nonumber\\[4mm]
&\times&  \int_0^{\infty}
rdr\left\{\left(\frac{d}{dr}\rho (r)\right)^2
+\frac1{r^2}f_3^2(1-\rho )^2\right.
\nonumber\\[4mm]
&+&
\left.
g_2^2\left[\frac12 \rho^2(\phi_1^2+\phi_2^2)
+(1-\rho )(\phi_1-\phi_2)^2\right]\right\}.
\label{anintZ}
\eeqn
Here we recognize the same normalization
integral which emerges in Sect.~\ref{kineticterm}. Keeping
in mind that $I=1$  we arrive at
\beq
Z_{4D} =-\frac{c}{2\pi}\, \int d z \, \pt_z\left(\chi_1^a\chi_2^a
-i\varepsilon^{abc}n^a\chi_1^b\chi_2^c\right)
\label{4danom}
\eeq
Rewriting the bifermion operator in (\ref{4danom}) in the holomorphic
representation \cite{NSVZ},
\beq
\chi_1^a\chi_2^a  -i\varepsilon^{abc}n^a\chi_1^b\chi_2^c=
-\frac{4}{(1+|w|^2)^2}\bar{\Psi}_{L}\Psi_{R}\,,
\label{chipsi}
\eeq
and  substituting it in (\ref{4danom}) we finally get
\beq
Z_{4D}=\frac{2c}{\pi}\int d z \pt_z
\left(\frac{1}{\zeta^2}\bar{\Psi}_{L}\Psi_{R}\right).
\label{4dcch}
\eeq

The right-hand side  of this equation reproduces the two-dimensional
anomaly (\ref{13thirteen}) provided that the coefficient
\beq
c=\frac 1{2\sqrt{2}}\,.
\label{otvet}
\eeq
The four-dimensional central charge (\ref{4dcch})
with the coefficient $c$ given by (\ref{otvet}) leads to the
result (\ref{mmmu0}) for the monopole mass in the limit $\Delta m=0$,
provided  we use the values of the fermion condensate (\ref{fercond}) in
the two vacua (at $z=\pm \infty$) of the $CP^1$ model.

\section{2D sigma-model kink and 4D
Seiberg-Witten exact solution}
\label{dopo6}
\setcounter{equation}{0}

Why  the 't Hooft-Polyakov  monopole mass   (i.e. on the Coulomb branch
at $\xi=0$) is given by the same formula (\ref{mmCb}) as  the mass
(\ref{mm})  of
the strongly confined  large-$\xi$ monopole
(subject to condition (\ref{lmxi}))? This fact was noted  in
Sect.~\ref{stringjunctionsolution}. Now we will explain the reason lying
behind this observation. {\em En route}, we will explain another striking
observation made in Ref.~\cite{Dorey}. A remarkably close parallel between
the
four-dimensional Yang-Mills theory with $N_f=2$ and the two-dimensional
$CP^1$ model was noted, at an observational level,
by virtue of comparison of  the corresponding central charges.
The observation was made on the Coulomb branch
of the Seiberg-Witten theory, with unconfined 't Hooft-Polyakov-like
monopoles/dyons. Valuable as it is,
the parallel was quite puzzling since the solution of the  $CP^1$ model
seemed to have no physics connection to the Seiberg-Witten solution. The
latter
gives the mass of the unconfined
monopole in the Coulomb phase  at $\xi=0$
while the $CP^1$ model emerges only in the   Higgs phase of the microscopic
theory.

Physics lying behind the above remarkable parallel will be revealed here.
First and foremost,  the previous study revealed the fact that the $CP^1$
model is a macroscopic description of the
four-dimensional Yang-Mills theory with $N_f=2$ in the Higgs phase.
This establishes a direct correspondence between the $CP^1$
model and our microscopic model at $| \mu | \ll \xi$.
Needless to say, the correspondence covers the central charges, CMS,
dyonic excitations, and {\em all other features} of the two theories
in 4D and 2D,
respectively.

Now we will show, that in the BPS sector
(and {\em only} in this sector) the correspondence extends further,
since the parameter $\xi$, in fact, cannot enter relevant formulae.
Therefore, one can vary $\xi$ at will, in particular,
making it less than  $| \mu | $ or even tending to zero,
where $CP^1$ is no more the macroscopic model for our microscopic theory.
Nevertheless, the parallel expressions for the central charges
and other BPS data in 4D and 2D, trivially
established at $| \mu | \ll \xi$, will continue to hold
even on the Coulomb branch.
The  ``strange coincidence" we observed in
Sect.~\ref{stringjunctionsolution}
is no accident. We deal here with
 an exact relation which stays valid including both
perturbative and nonperturbative corrections.

Physically the monopole in the Coulomb phase is very different from the one
in the confinement phase, see Fig.~\ref{twoabcd}. In the Coulomb phase it is
a 't Hooft-Polyakov monopole, while in the confinement phase it becomes
related to a junction of two non-Abelian strings. Still let us show
that the masses of these two objects are given by the same expression,
\beq
M_m^{\rm Coulomb }= M_m^{\rm confinement }\,,
\label{mmCc}
\eeq
provided that $\Delta m$ and the gauge couplings are kept fixed.
The superscripts refer to the Coulomb and monopole-confining phases,
respectively.

Our point is that the mass of the monopole cannot depend on the FI
parameter $\xi$. Start from the monopole in the Coulomb phase
at $\xi=0$. Its mass is given by  the exact Seiberg-Witten formula \cite{SW2}
\beq
M_m^{\rm Coulomb}=
\sqrt{2}\left| a_D^3\, \left(a^3=-\frac{\mu}{\sqrt{2}} \right) \right|
=\left|\frac{\mu}{\pi}\ln{\frac{\mu}{\Lambda}} +\mu\sum_{k=1}^{\infty}
c_k\left(\frac{\Lambda}{\mu}\right)^{2k}\right|\, ,
\label{mmSW}
\eeq
where $a_D^3$ is the dual Seiberg-Witten  potential for the SU(2)
gauge subgroup, and we take into account that for $N_f=2$ the first
coefficient
of the $\beta$ function is  2. Here $a^3=- {\mu}/\sqrt{2}$ is the
argument of
$a_D^3$,
the logarithmic term takes into account the one-loop result  (\ref{g2})
for the
SU(2) gauge coupling at the scale $\mu$, while the power
series is the expansion in instanton-induced corrections.

Now, if we introduce a small FI parameter $\xi\neq 0$ in the theory,
on dimensional grounds, we could expect in Eq.~(\ref{mmSW}) corrections to
the monopole mass  in powers of $\sqrt{\xi}/\Lambda$
and/or $\sqrt{\xi}/\mu$. These corrections are {\em forbidden}
by the U(1)$_R$ charges. Namely, the U(1)$_R$ charges of $\Lambda$ and
$\mu=\Delta m$ are equal to 2 (and so is the U(1)$_R$ charge of the
central charge under consideration)
while $\xi$ has a vanishing U(1)$_R$ charge.
For convenience, the U(1)$_R$ charges of different fields and parameters
of the microscopic theory are collected in Table~\ref{table3}. Thus, neither
$(\sqrt{\xi}/\Lambda )^k $ nor $(\sqrt{\xi}/\mu )^k$ can appear.

\begin{table}
\begin{center}
\begin{tabular}{|c|c | c| c|c| c | c | c| c|}
\hline
Field/parameter  & $a$ & $a^a$ & $\lambda^{\alpha}$ & $q$ &
$\psi^{\alpha}$ & $m_{A}$ & $\Lambda$ & $\xi$
\\[3mm]
\hline
U(1)$_R$ charge & 2 & 2 & 1 & 0 & $-1$
& 2 & 2 & 0
\\[2mm]
\hline
\end{tabular}
\end{center}
\caption{The U(1)$_R$
charges of fields and parameters of the microscopic theory.}
\label{table3}
\end{table}

By the same token, we could start from the confined monopole at large $\xi$,
and study the dependence of the monopole (string junction) mass
as a function of $\xi$ as we reduce $\xi$. Again, the above  arguments
based on the U(1)$_R$  charges tell us  that corrections in powers
of  $\Lambda/\sqrt{\xi}$ and $\mu/\sqrt{\xi}$  cannot appear.
This leads us to  Eq. (\ref{mmCc}).

Another way to arrive at the same conclusion is to observe that
the mass of the monopole is determined by the central charge
(\ref{4dcc}). This central charge is a holomorphic quantity
and, thus, cannot depend on the FI parameter $\xi$ which is not holomorphic
(it is a component of the SU(2)$_R$ triplet \cite{VY}).

Now recall that the mass of the monopole in the confinement phase
is given by the  kink mass   in the \ntwo $CP^1$ model, see
(\ref{monkink}). Thus, we obtain
\beq
M_m^{\rm Coulomb }\leftrightarrow M_m^{\rm confinement }
\leftrightarrow M_{\rm kink}\,.
\label{mmk}
\eeq
In particular, at the one-loop level, the kink
mass is  given by Eq. (\ref{exhau}). This leads to the relation
$\Lambda_{2D}=\Lambda$
between the 2D and 4D dynamical scales
which we noted earlier as a ``strange
coincidence," see Eq. (\ref{lambdasig}). Now we know the
physical reason behind it.
A puzzling question immediately coming to one's mind is
what would happen with more quark flavors.
One may suspect that  adding
extra quark flavors in  our microscopic theory (i.e. more than two)
will change its renormalization-group  flow
while  the renormalization-group  flow in the macroscopic model
seemingly  remains the same, which would certainly destroy
the correspondence.
This conclusion is wrong because, with more than two flavors in the
microscopic
theory,  the
strings that form in the microscopic
theory
become semilocal \cite{AV}. Semilocal strings has
additional zero modes associated with the change in their transverse
size. Thus, the moduli space of these strings changes, and is no longer
given by the $CP^1$ model \cite{Hanany:2003hp}. The macroscopic
model is just different.

Summarizing, the exact expression for the BPS kink masses in the 2D sigma
model is given by $|Z_{2D}|$, see Eqs.~(\ref{nickone}) and (\ref{nicktwo}).
These are also the expressions for the confined monopoles (and dyonic
monopoles)
in the 4D Yang-Mills theory, and, in view of the above,
the expressions coinciding with the SU(2) Seiberg-Witten monopole/dyon
solution on the Coulomb branch at the particular point
$ a^3=-\mu/\sqrt{2}$.  Although we do not discuss it
in the present paper, the above relation can be generalized (cf.~\cite{Dorey,HaHo}) to theories with the SU($N$)$\times$U(1)
gauge group and $N_f=N$  flavors on the four-dimensional
side, and $CP^{(N-1)}$ sigma models on the two-dimensional side.
This is because the effective world-sheet theory for non-Abelian
strings in \ntwo QCD with the
SU($N$)$\times$U(1) gauge group and $N_f=N$ flavors
is the $CP^{(N-1)}$ sigma model
\cite{Hanany:2003hp,Auzzi:2003fs}. We are planning to return to this
issue elsewhere.
%
%

\section{Conclusions}
\label{conclusions}
\setcounter{equation}{0}

In this paper we studied various dynamical regimes associated
with the confined monopoles that occur on the Higgs branch
of ${\cal N}=2$ two-flavor QCD. The focus of our consideration is
the quasiclassical treatment of the string junctions in the domain
$$\Lambda \ll |\Delta m |\ll \sqrt\xi\,.$$
The BPS sector is fully solvable. The confined monopole
carries two elementary strings attached to it, and can be viewed
as a string junction. We derived a complete set of the first-order
master equations, and found their solutions
corresponding to 1/2 and 1/4 BPS saturation. The string junction is
1/4 BPS. We obtained the orientational and superorientational
zero modes, introduced the corresponding moduli (quasi-moduli)
and developed a macroscopic description of strings and their junctions
based on two-dimensional $CP^1$ model with a twisted mass related
(equal) to the difference of the mass parameters of two flavors
in our  microscopic theory. The masses and other characteristics of
the confined monopoles are matched with those of the
$CP^1$-model kinks. The matching reveals, in particular,
the occurrence of an anomaly in the monopole central charge
in 4D Yang-Mills theory.

Building on  the established identification of
the microscopic and microscopic models, we expand in two crucial directions.
We study what becomes of the confined  monopole in the {\em bona fide}
non-Abelian limit $\Delta m \to 0$
where the global SU(2) symmetry is restored.
To this end we considered such monopoles (classically they would
become massless and infinitely spread)  as quantum states
interpolating between two vacua of the $CP^1$ model with vanishing
twisted mass.
This is a highly quantum regime, whose solution is known,
however, e.g. from the mirror description of the $CP^1$ model.
The  classical would-be explosion  never happens.
Instead, the monopole becomes stabilized by nonperturbative
dynamics  in the  effective 2D sigma model on the string world sheet.
This monopole, aka the $CP^1$-model kink, acquires a
nonvanishing  mass of the order of $\Lambda$ and a finite  size of the
order of
$\Lambda^{-1}$.

Another direction is the small-$\xi$ domain.
If $\Delta m $ is kept fixed, while $\xi$ decreases,
we move towards a weaker confinement, eventually ending
up on the Coulomb branch (i.e.  $\xi =0$)
where the Seiberg-Witten exact solution applies.
Needless to say, in this limit the $CP^1$ model is irrelevant
to the macroscopic description.  In the Coulomb phase
the SU(2) gauge subgroup  gets broken  down to U(1)
 \cite{SW1}, but the residual U(1)
is {\em not} broken if ${\cal N}=2$ is maintained.
The breaking of U(1) occurs only if an explicit
breaking of ${\cal N}=2$ is introduced ``by hand,"
and strings which develop in this case are Abelian (Abrikosov-Nielsen-Olesen) strings. The monopole never becomes {\em bona fide}
non-Abelian, although it acquires a mass given by the  Seiberg-Witten formula. Physics of the Coulomb phase has
nothing to do with that of the $CP^1$ model.
And, nevertheless, as was observed in Ref.~\cite{Dorey},
the BPS spectrum of the $CP^1$ model is in one-to-one
correspondence with the exact Seiberg-Witten solution.

The puzzle is solved by the following observation.
Physics of our microscopic model in  the Higgs phase, $\xi\neq 0$,  is
perfectly similar  to that of the $CP^1$ model. More precisely, the BPS
sectors can be mapped one onto another. On the other hand,
holomorphic nature of the central charges
preclude them from developing a $\xi$ dependence.
Therefore, BPS data in the Higgs phase are related to those in the
Coulomb phase.

Needless to say, of more practical interest is
condensation of non-Abelian mono\-poles which must be responsible for
 non-Abelian confinement of
quarks -- a phenomenon dual to the one studied in this paper,
where confined are non-Abelian monopoles, while quarks condense.
We are not sure how to dualize our results.
One way to approach this problem is a reduction of  the quark mass parameters $m_{1,2}$. If they get sufficiently small, below
the Argyres-Douglas point \cite{AD},  the quarks  acquire magnetic
quantum numbers \cite{BF} and their condensation should trigger  confinement of color-electric charges.

\section*{Acknowledgments}

We are grateful to  Alexander Gorsky, Adam Ritz and
Arkady Vainshtein for very useful discussions. We would like
to thank N. Dorey for extremely valuable communications.
The work  of M.S. is supported in part by DOE grant 
DE-FG02-94ER408. A.~Y. is supported in part by the Russian 
Foundation for Basic
Research   grant No.~02-02-17115, by INTAS grant
No.~00-00334 and by Theoretical Physics Institute
at the University of Minnesota.

\vspace{0.8cm}
\noindent
{\bf Note added (March 13, 2004)}

\vspace{0.2cm}
\noindent
David Tong has just informed us that he is finalizing a paper,
a sequel to his inspiring publication \cite{Tong:2003pz},
which has an overlap with the results reported here.

\vspace{0.8cm}
\noindent
{\bf Note added (May 22, 2004)}

\vspace{0.2cm}
\noindent
After this paper was submitted (and accepted)
for publication in Physical Review D we became aware of
some new circumstances. First, string junctions in the Yang-Mills-Higgs
system were considered (in a nonsupersymmetric context) in the 1980's.
In a model with the SU(2)$\to$U(1)$\to Z_2$ symmetry breaking pattern
it was found \cite{d-one} that there exist two distinct (degenerate) 
strings, each carrying $2\pi/e$ units of the magnetic flux.
Although these strings had no orientational moduli, they did support
kinks interpolating between them, which were called  ``monopole beads"
by the authors. Topological arguments  were presented \cite{d-one} 
proving the stability of such field configurations. Further analysis 
was carried out in Ref.~\cite{d-two} where it was shown that, in fact,  
many popular models of the type $G\to H\to Z_2$ do {\em not} support 
stable monopole beads. Non-translational zero modes of strings were discussed
in a U(1)$\times$U(1) model \cite{d-W,d-three}, and later, in more 
contrived models, in Ref.~\cite{d-four} (the latter paper is entitled 
``Zero Modes Of Non-Abelian Vortices").

It is worth emphasizing that, along with apparent similarities of which we 
will say later, there are drastic distinctions between the ``non-Abelian strings"
we consider here and the strings that were discussed in the 1980's.
In particular, in the example treated  in Ref.~\cite{d-four} 
the gauge group is not completely broken in the vacuum, and, therefore, there 
are massless gauge fields in the bulk. If the unbroken generator acts nontrivially
on the string flux (which is proportional to a broken generator)
then it can and does create zero modes. Some divergence problems ensue.

In contrast, in our case the gauge group is completely broken
(up to a discrete subgroup $Z_2$). The theory
in the bulk is fully Higgsed. The unbroken 
group SU(2)$_{C+F}$, a combination of the gauge and flavor groups, is global.
There are no massless fields in the bulk.

We could model the example considered 
in \cite{d-four} if we gauge our unbroken global symmetry 
SU(2)$_{C+F}$ with respect to yet another {\em ad hoc} gauge field $B_{\mu}$.

Some technical points first introduced in Ref.~\cite{d-four}
are close to constructions exploited in our papers 
\cite{Shifman:2002jm,SYsu3wall,SYmeta} and the present paper.
In particular, generation of $k=0,3$ components
of the gauge potential upon switching on the
$t,z$ dependence of the moduli, determined by an extra profile function
$\rho$ (see Eq. (4.1)) was first implemented in \cite{d-four}.

We are very grateful to Mark Hindmarsh for pointing out to us
the above publications which, unfortunately, escaped our attention.

We use this opportunity to add that the
Hanany-Tong paper mentioned in our Note of March 13, was posted on
March 15 \cite{HTadd}. Another relevant development was reported in a
publication of Isozumi {\em et al.} \cite{d-five} which was posted
on May 16. The authors obtained a general solution of 1/4 BPS equations
similar to those we have derived and discussed in the bulk of the paper (see also
our previous works \cite{Shifman:2002jm,SYsu3wall}).
Their construction is applicable in the strong coupling limit
in U($N)$ gauge theories with
the number of fundamental hypermultiplets exceeding $N$.
In fact, due to this reason, their strings are
``semilocal strings."  In this case one has
Higgs branches instead of isolated vacua, as is the case
in our analysis. When one goes to the strong coupling limit the
model effectively reduces to a sigma model on the Higgs branch.
Strings become instantons lifted to 4D (semilocal strings).
In this case solving the first order equations seems  to be an
easier task than in ours. This paves the way to a full  analytical
solution.

\newpage


\end{document}